\documentclass[a4paper,fleqn,usenatbib]{mnras}

\usepackage[T1]{fontenc}
\usepackage{ae,aecompl}

\usepackage{lscape}

\usepackage{graphicx}	
\usepackage{amsmath}	
\usepackage{amssymb}	
\usepackage{caption}
 
 \usepackage{soul}
 
\DeclareMathOperator{\sech}{sech}

\def\aj{AJ}%
\def\apj{ApJ}%
\def\apjl{ApJ}%
\def\apjs{ApJS}%
\def\ao{Appl.~Opt.}%
\def\aap{A\&A}%
\def\aapr{A\&A~Rev.}%
\def\mnras{MNRAS}%
\def\pasp{PASP}%
\def\pasj{PASJ}%
\def\bain{Bull.~Astron.~Inst.~Netherlands}%
\def\fcp{Fund.~Cosmic~Phys.}%
\begin{document}

\title[The Size and Shape of the Milky Way in M-dwarfs]{The Size and Shape of the Milky Way Disk and Halo from M-type Brown Dwarfs in the BoRG Survey}

\author[van Vledder et al.]{Isabel van Vledder $^{1}$\footnotemark[1],
Dieuwertje van der Vlugt$^{1}$\footnotemark[2], B.W. Holwerda$^{1}$\footnotemark[3]
\newauthor
M. A. Kenworthy$^{1}$, R. J. Bouwens$^{1}$, M. Trenti$^{2}$ \\
\footnotemark[1] Student ID: 1287257, E-mail: isamvv@yahoo.com\\
\footnotemark[2] Student ID: 1288024, E-mail: dieuwertjevdvlugt@hotmail.com\\
\footnotemark[3] Corresponding Author. E-mail: holwerda@strw.leidenuniv.nl, Twitter: @benneholwerda\\
$^{1}$ University of Leiden, Sterrenwacht Leiden, Niels Bohrweg 2, NL-2333 CA Leiden, The Netherlands\\
$^{2}$ School of Physics, University of Melbourne VIC 3010, Australia.}

\date{Accepted --- . Received --- ; in original form ---}

\pagerange{\pageref{firstpage}--\pageref{lastpage}} \pubyear{2015}

\maketitle

\label{firstpage}
\begin{abstract}
We have identified 274 M-type Brown Dwarfs in the Hubble Space Telescope's Wide Field Camera 3 (WFC3) pure parallel fields from the Brightest of Reionizing Galaxies (BoRG) survey for high redshift galaxies. These are near-infrared observations with multiple lines-of-sight out of our Milky Way.
Using these observed M-type Brown Dwarfs we fitted a Galactic disk and halo model with a Markov chain Monte Carlo (MCMC) analysis. This model worked best with the scale length of the disk fixed at $h$ = 2.6 kpc. For the scale height of the disk, we found $z_0 = 0.29^{+0.02}_{-0.019}$ kpc and for the central number density $\rho_0 = 0.29^{+0.20}_{-0.13}$ \#/pc$^3$. For the halo we derived a flattening parameter $\kappa$ = 0.45$\pm{0.04}$ and a power-law index $p$ = 2.4$\pm{0.07}$. We found the fraction of M-type brown dwarfs in the local density that belong to the halo to be $f_{h}$ = 0.0075$^{+0.0025}_{-0.0019}$. We found no correlation between subtype of M-dwarf and any model parameters.
{
The total number of M-type Brown Dwarfs in the disk and halo was determined to be $58.2^{+9.81}_{-6.70} \times10^{9}$. We found an upper limit for the fraction of M-type Brown Dwarfs in the halo of 7$^{+5}_{-4}$\%. The upper limit for the total Galactic Disk mass in M-dwarfs is $4.34^{+0.73}_{-0.5}\times10^{9}$ $M_{\odot}$,  assuming all M-type Brown Dwarfs have a mass of $80 M_J$.}
\end{abstract}


\section{Introduction}
\label{s:intro}

Counting stars in our Galaxy has long been used to infer its structure. This is mostly done with relatively luminous stars due to insufficient data on substellar objects. Early work has been done by \cite{Kapteyn22}, \cite{Seares25} and \cite{Oort38} who used this method to determine the geometrical structure of the Galaxy.

In this paper, the focus lies on the M-type brown dwarfs (hereafter M-dwarfs). Brown dwarfs are very dim (sub)stellar objects with masses that range from $13 M_J$ to $80 M_J$. They are not able to fuse hydrogen and thus are not considered stars. Instead, they burn deuterium and lithium. The lower limit of burning deuterium is $13 M_J$ and that of lithium burning is around $60 M_J$. Brown dwarfs have a limited amount of nuclear energy because of the exothermic reactions of deuterium and lithium, making them cool over time.  M-dwarfs are the hottest of their kind followed by L-, T-, and Y-dwarfs \citep{LeBlanc}. These types are divided in subtypes where 0 indicates the hottest and 9 the coolest of a particular type. M0 objects are not classified as brown dwarfs, but as low-mass stars. However, because they are dim low mass objects with an M-type colors, we will include them in this paper. Brown dwarfs are believed to be among the most numerous luminous objects in our Milky Way. Studying them can thus tell us a lot about the structure of the Milky Way.  

Brown dwarfs resemble high redshift galaxies in both colour and angular size. For example, redshift $z\sim7$ galaxies have very similar broad-band colors as L-dwarfs and both are unresolved in most ground-based images. Most of the time, we are still able to distinguish between them because of their different sizes in HST imaging (stars remain unresolved with $FWHM < 0\farcs1$). High redshift galaxies usually appear fuzzier than brown dwarfs making their FWHM larger. This is not the case with wide-field imaging from the ground, $z$ $>$ 5 galaxies are then unresolved at the seeing limit \citep{Stanway08}. At faint magnitudes, it becomes hard to resolve galaxies in order to separate them from brown dwarfs using high angular-resolution imaging \citep{Tilvi13}. Brown dwarfs and high redshift galaxies can therefore easily be confused with each other. Thus, for many surveys looking for $z$ $>$ 5 galaxies, brown dwarfs remain the main contaminants. 

Where morphological information is not available, it is still possible to identify M- and L-dwarfs on their red colours in near infrared \citep{Stanway08}. They can be distinguished from $z$ $\sim$ 5-6 galaxies on the basis of their spectra. Like Lyman-break galaxies M- and L-dwarfs show abrupt breaks in their spectra, but deep molecular absorption lines in the continuum longwards of the first detected break allow observers to distinguish them from high redshift galaxies. However, spectroscopy has proven to be challenging and observationally expensive. Spectrographs cannot reach the continuum level for dim sources with typical magnitudes of $J_{AB}$ $\sim$ 27.5 \citep{Wilkins14}. A good understanding of the initial mass function (IMF) is also needed, especially at the low mass end because the IMF can be used to estimate the fraction of brown dwarfs in surveys.

\begin{figure*}
\centering
\includegraphics[width=\linewidth]{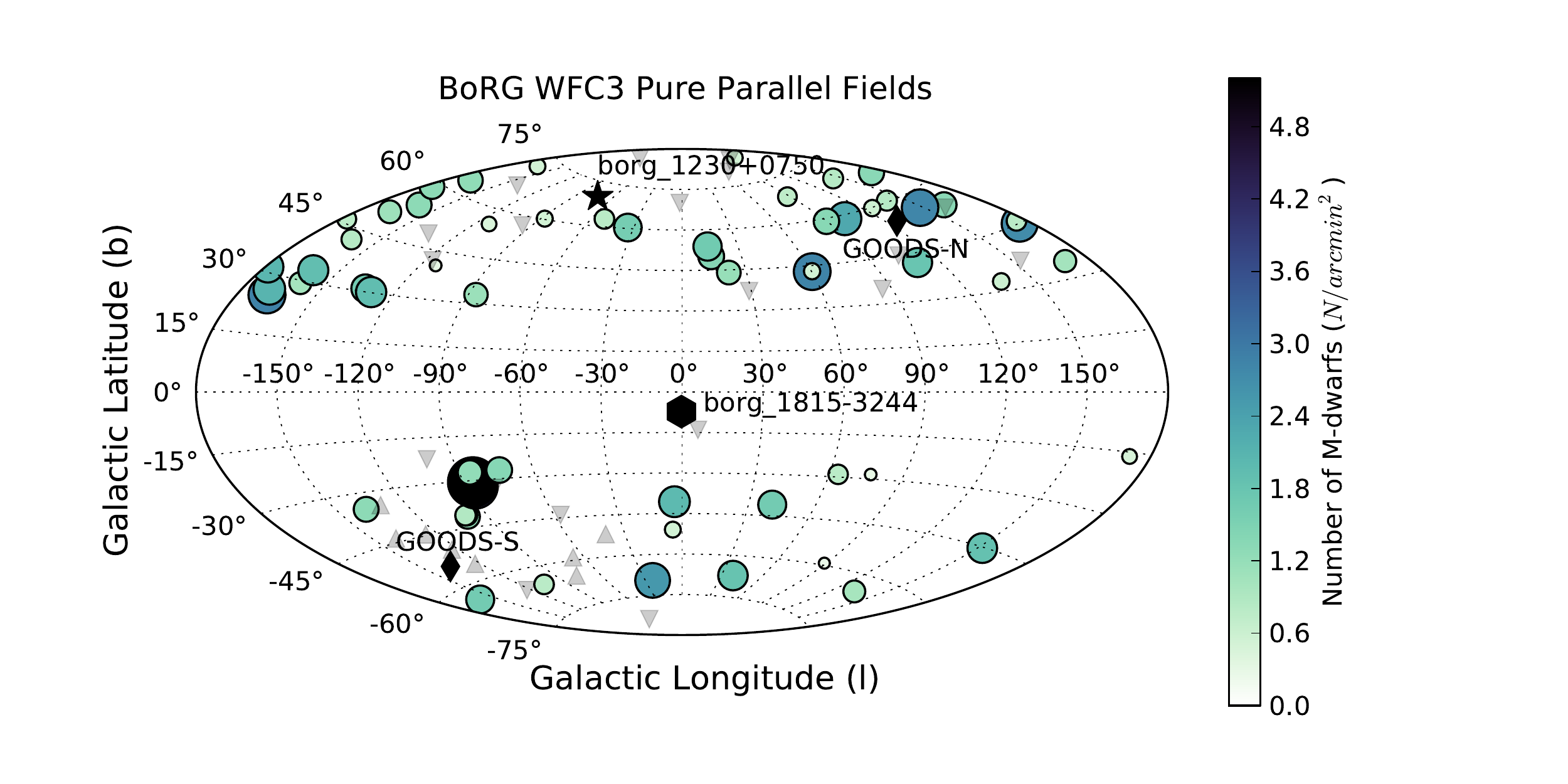}
\caption{{
Distribution of BoRG fields and satellite galaxies with the number of M-dwarfs indicated as both the color and size of the symbol. The fields that are discarded are also indicated; Sagittarius stream field (star) and bulge field (hexagon). Neither of these contain $z\sim8$ galaxies. The grey triangles indicate the satellite galaxies.}}
\label{fig:Distribution}
\end{figure*}

Several authors have used the small numbers of stars in deep Hubble observations to determine the distribution of low-mass stars in the Milky Way.
For instance, \cite{Pirzkal05} determined the scale-height of different types of dwarfs from the Hubble Ultra Deep Field \citep[HUDF,][]{Beckwith06}. 
\cite{Ryan05} found L and T dwarfs in a small set of ACS parallels. \cite{Stanway08} and \cite{Pirzkal09} determined the Galactic scale-height of M dwarfs from the Great Observatories Origins Deep Survey fields \citep[GOODS,][]{goods}. These studies gradually improved statistics on distant L, T and M dwarfs to several dozen objects. 
The number of known dwarfs increased once again with the WFC3 pure-parallel searches for z$\sim$8 galaxies \citep[][]{Ryan11, Holwerda14}, 
taking advantage of many new sightlines (Figure \ref{fig:Distribution}). However, these studies are limited to the local scale-height of the dwarfs in the MW disk 
as the original observations are extra-Galactic and therefore avoid the plane of the Galaxy.

The IMF is a distribution of stellar and substellar masses in galaxies when they start to form. From the mass of a star, its structure and evolution can be inferred. Likewise, knowing the IMF is a very important step in understanding theories on star formation in galaxies. It can be seen as the link between stellar and galactic evolution \citep{Scalo}. 

The integrated galactic initial mass function (IGIMF) gives the total stellar mass function of all stars in a galaxy. It is the sum of all star formation events in the galaxy which would be correct in any case in contrast to a galaxy wide IMF, which was derived from star cluster scales \citep{Weidner13}. As the distribution of stellar masses has a big impact on many aspects of the evolution of galaxies, it is important to know to what extent the IGIMF deviates from the underlying stellar IMF \citep{Haas}.

In this paper, we derive the number of M-dwarfs in our Galaxy, which can be helpful for ultimately determining the IMF and the IGIMF. One can also estimate the amount of contamination in surveys of high redshift galaxies if morphological information is not available. However, the primary goal of this study is to find the number of M-dwarfs in our Milky Way galaxy and to learn more about its shape. For this we will use a model of the exponential disk \citep{van-der-Kruit81} combined with a power-law halo \citep{Chang} (Section \ref{s:3dmodel}).

The fit with the model of the exponential disk has been done before by \cite{Juric08}. They made use of the Sloan Digital Sky Survey (SDSS), which has a distance range from 100 pc to 20 kpc and covers 6500 $deg^2$ of the sky. They find that the number density distribution of stars as traced by M-dwarfs in the solar neighbourhood ($D < 2$ kpc) can best be described as having a thin and thick disk. They estimate a scale height and scale length of the thin and thick disk of respectively $z_0 = 0.3\pm{0.06}$ kpc and $h = 2.6\pm{0.52}$ kpc, and $z_{0} = 0.9\pm{0.18}$ kpc and $h = 3.6\pm{0.72}$ kpc. In the same way \cite{Pirzkal09} derived a scale height of $z_0 = 0.3\pm{0.07}$ kpc for the thin disk, but they made use of spectroscopically identified dwarfs with spectral type M0-M9.

\cite{Juric08} have also fitted the halo and found for the flattening parameter $\kappa = 0.64\pm{0.13}$, for the power-law index $p = 2.8\pm{0.56}$ and for the fraction of halo stars in the local density $f_{h} = 0.005\pm{0.001}$. Similar results were obtained by \cite{Chang}. They made use of the $K_{s}$-band star count of the Two Micron All Sky Survey (2MASS) and found $\kappa = 0.55\pm{0.15}$, $p = 2.6\pm{0.6}$ and $f_{h} = 0.002\pm{0.001}$.

In this paper, however, we do not assume the exponential disk consists of two separate components, a distinct thin and thick disk, like \cite{Juric08} and \cite{Chang} do (see section \ref{s:3dmodel} for a further discussion) but we treat the disk as a single component. This has also been done by \cite{Holwerda14} for a disk-only fit. 
{ We make use of a Python implementation of Goodman and Weare's Markov Chain Monte Carlo (MCMC) Ensemble sampler called }$emcee$ \citep{MCMC} { and include a Galactic halo contribution.}

This paper is organized as follows: \S \ref{s:data} describes our starting data and how M-dwarfs were identified, \S \ref{s:3dmodel} describes the three models we use in MCMC to describe the distribution of M-dwarfs in the BoRG data, \S \ref{s:MCMCfit} describes out implementation of the MCMC fit to this problem, \S \ref{s:analysis} describes the analysis of the three MCMC models in detail and the implied number of M-dwarfs, \S \ref{s:discussion} is our discussion of the results. \S \ref{s:conclusions} lists our conclusions and \S \ref{s:future} outlined future options for the discovery and modeling of M-dwarfs with e.g., EUCLID or WFIRST. 
%
%

\begin{figure}
\centering
\includegraphics[width=\linewidth]{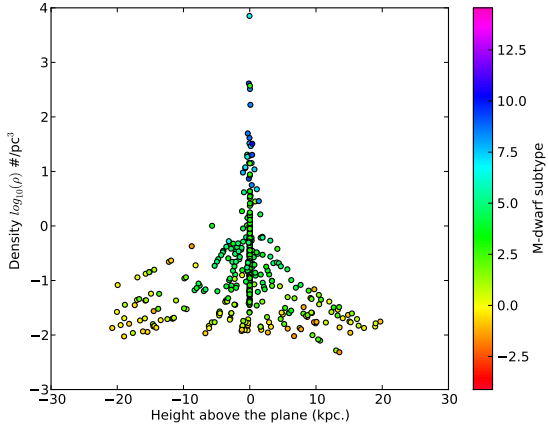}
\caption{The inferred height above the plane of the disk from the distance modulus and the new Galactic Coordinates, regardless of radial position, with the M-dwarf photometric subtype marked (color bar). }
\label{fig:rawheights}
\end{figure}

\section{Data}
\label{s:data}

For the model fits we use data similar to Table 14 in \cite{Holwerda14} and new reduced data acquired from the BoRG survey \citep{Trenti11}. This is a pure-parrallel program with the Hubble Space Telescope using the Wide Field Camera 3 (WFC3). The survey targets the brightest galaxies at $z\sim8$, suspected to be the prime source of reionizing photons in the early Universe \citep{Bradley12,Schmidt14}. The strategy of four filters, three near-infrared ones and a single optical one, has now been shown to work extremely well to identify and approximate type brown dwarfs \citep{Ryan11,Holwerda14}. The pure-parallel nature of the program ensures random sampling of sky, lowering cosmic variance errors.

\subsection{New Galactic Coordinates}
\label{s:coords}

The data table in \cite{Holwerda14} contains an inadvertent problematic error: the Galactic coordinates are not correctly computed and thus the height above the plane and the galactic radius for all dwarfs are not correct. We traced this error back to an incorrect coordinate transformation in the package {\sc PyEphem} \citep{pyephem} where the equatorial coordinates were not correctly transformed into galactic coordinates. 
The galactic coordinates were recalculated with the package {\sc astropy}  \citep{Astropy}. The new values for the galactocentric radius and height above the plane were calculated with equations \ref{radius} and \ref{height}.

\begin{equation}\label{radius}
R = R_{\odot}^2 + d^2\cos(b)^2-2R_{\odot}d\,\cos(l)\cos(b),
\end{equation}
\begin{equation}\label{height}
z = d\,\sin(b)^2 + z_{\odot},
\end{equation}

\noindent where $R_{\odot}$ and $z_{\odot}$ are the position of the Sun respectively 8.5 kpc and 0.027 kpc \citep{Chen01}. $l$ is the galactic longitude and $b$ the galactic latitude. The new coordinates calculated with these equations and presented in Table \ref{t:Mdwarfs} and in Figure \ref{fig:rawheights}, together with additional M-dwarfs identified in additional BoRG fields. 
Most of the brown dwarfs found with the BoRG survey are not positioned in the disk but in the halo.  

\begin{table}
\begin{tabular}{|l|r|r|}
\hline
Type & Vega Magnitude & AB Magnitude\\
\hline
\hline
0 & 6.45 & 7.34\\
1 & 6.72 & 7.61\\
2 & 6.98 & 7.87\\
3 & 7.24 & 8.13\\
4 & 8.34 & 9.23\\
5 & 9.44 & 10.33\\
6 & 10.18 & 11.07\\
7 & 10.92 & 11.81\\
8 & 11.14 & 12.03\\
9 & 11.43 & 12.32\\
\hline
\hline
\end{tabular}
\caption{\label{tab:absolute magnitude} Absolute Vega and AB magnitudes in the 2MASS J-band (F125W) of
M-dwarfs from \protect\cite{Hawley02}.}
\label{t:hawley}
\end{table}

\subsection{ Identifying Brown Dwarfs}
\label{s:identifying brown dwarfs}

The brown dwarfs used in this research were found with the BoRG survey. Observations were made with the Wide Field Camera 3 (WFC3) aboard the Hubble Space Telescope (HST) during a pure parallel program. The WFC3 was taking exposures whilst the HST was pointing for primary spectroscopic observations on e.g., quasars (Figure \ref{fig:Distribution}). The near-random pointing nature of the program makes sure the BoRG fields are minimally affected by field-to-field (cosmic) variance \citep{Trenti08} and is therefore ideal to find the density distribution of M-dwarfs. The survey is designed to identify high redshift galaxies and uses four different filters: $F098W$, $F125W$, $F160W$ and $F606W$. The $F098W$ filter is designed to select the redshift $z \sim$ 7.5 galaxies (Y-band dropouts), $F125W$ and $F160W$ are used for source detection and characterization, and $F606W$ is used to control contamination from low redshift galaxies, AGN's and cool Milky Way stars like brown dwarfs. 

The brown dwarfs in the BoRG fields were identified from their morphology and colour. Using the bona-fide M-dwarf catalog for the CANDELS survey, three morphological selection criteria are defined in \cite{Holwerda14}: the half-light parameter, the flux ratio between two predefined apertures (stellarity index) and the relation between the brightest pixel surface brightness and total source luminosity (mu\_max/mag\_auto ratio). The selection criteria are defined such that they picked out the 24 by the PEARS project spectroscopically identified M-dwarfs \citep{Pirzkal09}. The half-light parameter and the mu\_max/mag\_auto worked well for stellar selection brighter than 24 mag (adopted here), and the half-light radius includes much fewer interlopers down to 25.5 mag. The half-light parameter and the stellarity selection criteria seemed to be the most appropriate for the BoRG fields because the locus of stellar points turned out to be within the criteria lines. The mu\_max/mag\_auto criterion did not work as well because this criterion is sensitive to pixel size and the CANDELS data was originally at a different pixel size. Therefore, the half-light parameter is used as the morphological selection criterion.

The various spectral types (M, L, T) are identified by construction of a $J_{F125W} - H_{F160W}$ vs. $Y_{F098M} - J_{F125W}$ near-infrared colour-colour diagram. The colour-colour criterion to select M-dwarfs is based on the distribution of the PEARS-identified M-dwarfs in the CANDELS and ERS catalog. The colour-colour criterion to select M-, T- and L-dwarfs is drawn from \cite{Ryan11}. 

To find the subtypes of the found M-type dwarfs a linear relation is fitted to PEARS-identified M-dwarfs in CANDELS. This fit can be found in Figure 14 in \cite{Holwerda14}. The linear relation is expressed by Equation \ref{linear relation}.
\begin{equation}\label{linear relation}
	Mtype = 3.39\times[V_{F606W}-J_{F125W}]-3.78
\end{equation}
\noindent which directly implies a distance:
\begin{equation}\label{distance}
d = 10^{\frac{m-M}{5}+1},
\end{equation}
with $m$ the apparent magnitude and $M$ the absolute magnitude. We note that BoRG photometry is already corrected for Galactic extinction. In principle, the Galactic reddening is an upper limit to the amount of extinction by dust in the case of Galactic objects such as these M-dwarfs. However, with these high-lattitude fields, the difference would be minimal compared to the photometric error.
 The absolute magnitude is correlated with subtype, this correlation was found by \cite{Hawley02} and is given in Table \ref{t:hawley}. The apparent magnitude is measured and given in Table 14 in \cite{Holwerda14} and our Table \ref{t:Mdwarfs}.

\subsection{Data Limitations}
\label{s:datalimits}

There is still the possibility of contamination by M-type giants, other type subdwarfs, and AGN's in the data set but this is considered to be small. For example, the on-sky density of M-giants is $4.3 \times 10^{-5}$ M-giants/arcmin$^2$ \citep{Holwerda14,Bochanski14} which makes it unlikely that one is included in the data set. A similar argument applies to nearby other type subdwarfs, the volume probed at close distances is very small. We are confident that the morphological selection, luminosity limit, and the colour-colour restriction select a very clean sample of Milky Way M-dwarfs.

The saturation limit for the BoRG sextractor catalogs was kept at 50000 ADU, corresponding to a bright limiting magnitude of $\sim$6.6 AB mag. This places no upper brightness (i.e. lower distance) limit on the M-dwarf catalog. Unlike STIS or ACS-SBC, the WFC3/IR channel has no official object brightness limits and none were implemented for the BoRG observations.

\begin{figure}
\begin{center}
	\includegraphics[width=0.5\textwidth]{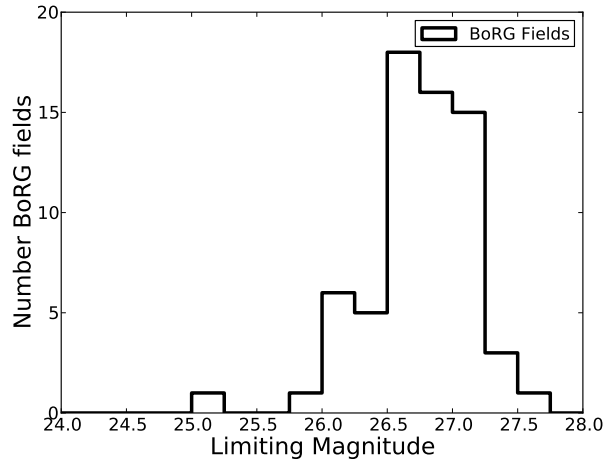}
	\caption{The {\em F125W}-band magnitude limits for the BoRG fields (from \protect\cite{Bradley12} we consider in the present study.}
	\label{f:limmag}
\end{center}
\end{figure}

{The BoRG survey's detection limits in J-band (F125W, varying from 26 to 27.5, form field-to-field, Figure} \ref{f:limmag}) {immediately inform us that M0 type brown dwarfs will be detected in the largest volume while the latest M-types in the smallest. Assuming an $m_{lim}\sim27$, this implies detection limits ranging from $\sim8$ kpc (M9) to $85$ kpc (M0) (See also Table} \ref{t:hawley}). {We note that these limiting distances are well into the Galactic Halo. For the shallowest field ($m_{lim} = 25$ AB), this translates to 3.5 (M9) and 35 (M0), still well out of the disk and into the Halo. The majority of our data is for M0/1 type M-dwarfs, for which the BoRG survey effectively samples up to $\sim30$ kpc or well into the Halo.}

{Unless they are in the Solar neighborhood, any binaries in the BoRG field catalog will be single star entries. For example, }\cite{Aberasturi14} {estimate the binary fractions of nearby ($<20$ pc) T-dwarfs using WFC3 at $\sim$16-24\% and they discuss how this confirms lower binary fractions with lower mass primaries. For M-dwarf primaries, the binary fraction is closer to 30-50\%} \citep[e.g.,][]{Janson12}. {That binary fraction is derived for nearby M-dwarfs in the disk and the fraction in the halo may be much lower (e.g. as the product of multiple star system ejection mechanisms). At a fiducial distance of 10 kpc, the $0\farcs1$ PSF of WFC3, would only be able to resolve a binary with 1 kAU separation. Therefore, we assume that each star identified here is a single M-dwarf. The factor 2 in flux (0.2 mag) is within the typical uncertainty of the} {\sc sextractor} {photometry. This bias is somewhat accounted for in the relatively low fiducial value we have given the data ($f$), i.e. individual data-points are noisy.} 

\subsection{Local Overdensities}
\label{s:overdensity}

The model used in this paper (Section \ref{s:3dmodel}) are for smooth stellar distributions. Substructures such as spiral arms, stellar streams and satellites, are not accounted for. To include these structures we need a model with many more parameters. The fitting of such a model lies beyond the scope of this data and this paper. Instead, we exclude fields in which we suspect these kinds of contamination and fit the remaining fields. We look for fields that show an strong overdensity and reject them based on their positions.

Two fields in the data set contain clear overdensities: borg\textunderscore$1230+0750$ and borg\textunderscore$1815-3244$. The overdensity of the first can be explained by the fact that its position is exactly on the Sagittarius stellar stream \citep{Majewski03,Belokurov06,Holwerda14}. The 22 M-dwarfs found in the field are therefore discarded for our analysis. We discard borg\textunderscore$1815-3244$ because of its low Galactic latitude and it is close to the plane of the disk. As a result of its position, the field is  vulnerable to contamination.

Another overdensity is positioned at Galactic latitude -30$^\circ$ \,and Galactic longitude -90$^\circ$. It contains 5 M-dwarfs which is high in comparison with the other fields. While keeping in mind that one of the eight newly discovered satellite galaxies, found by \cite{DES}, was close to this position, we plot the positions of those satellite galaxies to see if they match the position of the overdensity. We also plot the already known satellite galaxies to see if there is any overlap (Figure \ref{fig:Distribution}). The fields borg\textunderscore$0436-5259$, borg\textunderscore$0439-5317$ and borg\textunderscore$0440-5244$ are positioned around one of the known satellite galaxies. We also see that borg\textunderscore$1031+5052$ and borg\textunderscore$1033+5051$ are close to the galaxies found by the DES collaboration (Figure \ref{fig:Distribution}). In most cases the positions of the galaxies differ too much from the positions of the fields, about 2$^\circ$, for the galaxies to be the cause of any overdensities. We therefore include these fields in our analysis.
Out of the 72 fields of the BoRG2013 sample, we exclude two for obvious overdensities.

\section{A 3-D model of the Milky Way Disk}
\label{s:3dmodel}

The Milky Way Galaxy can be divided into four different components: the bulge, the halo, the thin disk and the thick disk, although the existence of distinct disks is sometimes questioned \citep[e.g.,][]{Bovy12b}.
The halo is built up from the stellar halo and the dark matter halo. The stellar halo contains about 2-10\% of the stellar mass in the Galaxy, mostly old stars with low metallicity. 
The bulge is a stellar system located in the center. It is thicker than the disk and it contains about 15\% of the total luminosity of the Galaxy. The stars in the bulge are believed to date from the beginning of the Galaxy. Because of lack of coverage on the Bulge, we discount this component in the following analysis.

The thick disk was discovered by star counts \citep{Yoshii82, Gilmore83} and contains stars that are older and have different composition from those in the thin disk. The thick disk is believed to be created when the infant thin disk encountered a smaller galaxy and the young disk was heated kinematically \citep{BT1}.
The thin and thick disk were believed to be distinct but recent research questions this. \cite{Bovy12b} examined the [$\alpha$/Fe] ratio, which is a proxy for age, of stars and their distribution. They found that old stars are distributed in disks with a small scale length and a great scale height and that, with decreasing age, the stars are distributed in disks with increasing scale length and decreasing scale height. Similar results were found by \cite{Cheng12} and \cite{Bensby11}. In addition to this, \cite{Bovy12b} found a smoothly decreasing function approximately $\Sigma_{R}(h) \propto exp(-h)$ for the surface-mass contributions of stellar populations with scale height $h$. This would not be expected if there was a clear distinction between the thick and the thin disk.
In addition, \cite{Chang} tried to fit different models for the thin and thick disk and found a degeneracy between all the parameters of those thin and thick disk models. 
We found a similar degeneracy in our model only using a disk model. 
They made use of the 2MASS catalog, which is an all sky survey. This sample should have been wide and deep enough to break degeneracies \citep{Juric08}. It seems therefore not useful to use two different models for the thin and thick disk. In this research we assume one model for the disk, presented in the next section.

\subsection{Galactic Disk Model}
\label{s:model for the disk}

For the disk we assume the following shape, which was found studying the three-dimensional light distribution in galactic disks \citep{van-der-Kruit81}:
\begin{equation}\label{model}
	\rho(R, z) = \rho_0\,e^{-\frac{R}{h}}\sech^2\left(\frac{z}{z_0}\right),
\end{equation}
\noindent where $\rho(R, z)$ is the dwarf number density in a point in the disk, $\rho_0$ the central number density, $R$ the galactocentric radius, $h$ the scale length, $z$ the height above the plane and $z_0$ the scale height of the disk. This is the first model we fit to the numbers of M-dwarfs.
To remain consistent with other exponential fits to the Milky Way disk, we report the fitted values for $\rho_0/4$ and $z_0/2$ \citep[see also][]{vdKruit81a}

\subsection{Galactic Halo Model}
\label{s:model for the halo}

The model used for fitting the halo (Equation \ref{halo model}) is based on the model of \cite{Chang} and contains a normalization for the position of the sun:

\begin{equation}\label{halo model}
\rho(R, z) = \rho_{\odot}\,f_{h}\,\left(\frac{R^2+(z/\kappa)^2}{R_{\odot}^2+(z_{\odot}/\kappa)^2}\right)^{-p/2},
\end{equation}
\noindent where  $\rho(R, z)$ is the dwarf number density at a point in the halo, $R$ the galactocentric radius and $z$ the height above the plane. ($R_{\odot}, z_{\odot}$) is the position of the Sun: (8.5 kpc, 0.027 kpc). $\rho_{\odot}(R_{\odot}, z_{\odot})$ is the local density, which is the density within a radius of 20 pc of the sun. This was found by \cite{Reid08}.
$f_{h}$ represents the fraction of stars in the local density that belong to the halo. The combination of the fraction, local density and normalization for the position of the sun can be seen as the central number density. $\kappa$ is the flattening parameter and $p$ is the power-law index of the halo.

The halo model has different parameters from the exponential disk model. The flattening parameter $\kappa$ is a measure for the compression of a sphere. It is defined as $\kappa = \dfrac{a-b}{a}$ with $a$ the semi-major axis and $b$ semi-minor axis. Because of this definition of $\kappa$, it must be between 0 and 1. 
The power-law index of the halo $p$ is the other new fit parameter and a positive number, due to the finite extent of the Milky Way \citep{Helmi}. 

\subsection{Galactic Disk+Halo Model}
\label{s:model for the halo and disk}

For the fit with the halo and disk a combination of Equations \ref{model} and \ref{halo model} is used: 
\begin{equation}\label{halo and disk model}
\rho(R, z) = \rho_0\,e^{-\frac{R}{h}}\sech^2\left(\frac{z}{z_0}\right) + \rho_{\odot}\,f_{h}\,\left(\frac{R^2+(z/\kappa)^2}{R_{\odot}^2+(z_{\odot}/\kappa)^2}\right)^{-p/2}
\end{equation}
\noindent which is the model we fit to the numbers of M-dwarfs in section \ref{s:halo and disk}.

\section{MCMC Fit}
\label{s:MCMCfit}

To find the best fitting model we use Bayesian analysis, which is a standard procedure in astronomy when measurement results are compared to predictions of a parameter-dependent model. 

\subsection{Bayesian Analysis}
\label{s:bayesian analysis}

For the Bayesian analysis we are using Bayes' theorem:

\begin{equation}\label{Bayes}
	\mathcal{P}(\theta|y,x, \sigma) = \frac{\mathcal{P}(y\,|\,x, \theta, \sigma,f)\mathcal{P}(\theta)}{\mathcal{P}(y\,|\,x, \sigma)},
\end{equation}
with $\mathcal{P}(\theta|y, x, \sigma)$ the posterior distribution, $\mathcal{P}(y\,|\,x, \theta, \sigma,f)$ the likelihood of $y$ given ($x$, $\theta$, $\sigma$, f) and $\mathcal{P}(\theta)$ the prior probability density. The prior probability density needs to be defined for the model parameters. This definition is based on previous research and observational data. $\mathcal{P}(y\,|\,x, \sigma)$ is the normalization constant which we assume is a constant because we are taking this ratio for the same physical model (see Section \ref{s:mcmc}). The posterior distribution is educed to:
\begin{equation}\label{posterior}
	\mathcal{P}(\theta|y, x, \sigma) \propto \mathcal{P}(y\,|\,x, \theta, \sigma,f)\mathcal{P}(\theta)
\end{equation}
To find an estimate for the parameters we need to marginalize the posterior distribution over nuisance parameters. This can by done with MCMC. An example of marginalization is shown in equation \ref{marginalization} where $\mathcal{P}(\theta_1|d)$ is the marginalized posterior distribution for the parameter $\theta_1$ \citep{Trotta}:
\begin{equation}\label{marginalization}
	\mathcal{P}(\theta_1|d) = \int \mathcal{P}(\theta|d)\mathrm{d}\theta_2\dotso \mathrm{d}\theta_n
\end{equation}

\subsubsection{Priors}
\label{s:priors} 

The MCMC implementation allows one to set priors for each of the variables; the ranges of plausible values for each variable. We set them such that
they exclude unphysical scenarios, negative densities, but these priors are not a hard top-hat; instead, their probability is set as very low.
In the case of an unphysical model, the MCMC model can in fact iterate towards an unphysical solution.  Such unphysical parameter values,
possibly in combination with highly quantized posterior (a poorly mixed MCMC chain) are one way to identify a poor model.

\subsubsection{Advantage of Bayesian Analysis}
\label{s:advantage} 

The main advantage of Bayesian analysis is that the method gives a basis for quantifying uncertainties in model parameters based on observations. The Bayesian posterior probability distribution depends on observations and the prior knowledge of the model parameters. 

The most criticized aspect of the Bayesian analysis is the necessity of defining the prior probability density. This prior can often be well estimated with the available data. In cases where this estimation is difficult the posterior distribution can be significantly influenced by the choice of the prior probability density. These kind of problems can also be found in frequentist methods where the choice of model parameters influences the results \citep{Ford}.

\subsection{MCMC Implementation}
\label{s:mcmc} 

For this research we use $emcee$ \citep{MCMC}, a Python implementation of the Markov chain Monte Carlo (MCMC) Ensemble sampler, to fit the model in Equation \ref{halo and disk model} to the data, with the Metropolis-Hastings method (We fit \ref{model} and \ref{halo model} as well with nonphysical results). MCMC provides us with an efficient way of solving the multidimensional integrals that we saw in the Bayesian analysis of models with many parameters. For a more in-depth explanation of MCMC, we refer the reader to \cite{MCMC}.

%

\subsection{Fit Variables}
\label{s:fit} 

We use the likelihood for $\mathcal{P}(y\,|\,x, \theta, \sigma,f)$ (Equation \ref{marginalization}) over all stars:  
\begin{equation}
\label{lnlike}
  \ln\mathcal{P}\,(y\,|\,x, \theta, \sigma,f) =
-\frac{1}{2} \sum_n \left[
    \frac{(y_n-model)^2}{s_n^2}
    + \ln \left ( 2\pi\,s_n^2 \right )
\right]
\end{equation}
In our case $y$ is the density of brown dwarfs $\rho(R, z)$, $x$ is given by $R$ and $z$, $model$ is given by Equation \ref{halo and disk model} for each $n$th star and $\theta$ is the set of free parameters for the model. For $s_n$ we now have:
\begin{equation}\label{sn}
s_n^2 = \sigma_n^2+f^2\,(model)^2,
\end{equation}
$f^2\,(model)^2$ contains the jitter which consists of all the noise not included in the measurement noise estimation $\sigma_n$, the error on the density of brown dwarfs $\rho(R, z)$ \citep{Hou12}. So $f$ gives the fraction of bad data and is a free parameter in the model. Marginalizing $f$ has the desirable effect of treating anything in the data that cannot be explained by the model as noise, leading to the most conservative estimates of the parameters \citep{Gregory05}.

\begin{figure}
\centering
\includegraphics[width=0.8\linewidth]{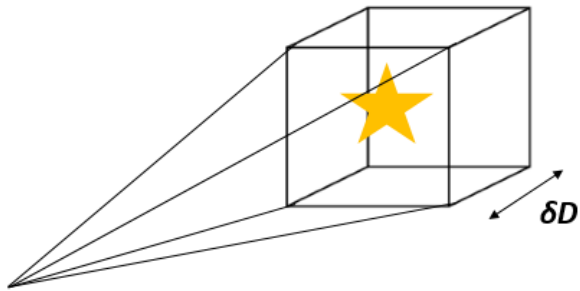}
\caption{The conversion of surface density to volume density with $\delta{D}$, one of the free parameters in our MCMC fit. }
\label{fig:deltaD}
\end{figure}

To compute the volume density $\rho(R, z)$ at each position of M-dwarfs, we first calculate the physical area at the inferred distance of the BoRG survey field in which the dwarf was found; the surface density. Then we multiply it with a bin-width $\delta{D}$ to get the volume density (Figure \ref{fig:deltaD}). This bin-width is a free parameter in the model. The physical area is calculated with the length of the field. This length is given by Equation \ref{length} with the typical size of the usable area of the observed BoRG field ($A_{BoRG}$ in arcminutes, typically about a WFC3 field-of-view).
The volume density is now given by Equation \ref{rho}.
\begin{equation}\label{length}
L = 2\tan(A_{BoRG})\times d
\end{equation}
\begin{equation}\label{rho}
\rho = \frac{1}{L^2\times\delta{D}}
\end{equation}
Because $\delta{D}$ is a free parameter the models we are going to fit, will now look like Equation \ref{model1}.
\begin{equation}\label{model1}
	\frac{1}{L^2} = \delta{D}\times\,model
\end{equation}

\section{Analysis}
\label{s:analysis}

To find the best fit of the data for our model, we need to find the numerical optimum of the log-likelihood function (Equation \ref{lnlike}) to get a starting position for MCMC. For this we use the module scipy.optimize. This minimizes functions, whereas we would like to find the maximum of the log-likelihood function. So with this module we use the negative log-likelihood function, which achieves the same. In doing so, we find the best starting values of the free parameters. The optimization module makes use of true parameters which are an initial guess for the parameters. We choose the Nelder-Mead method because of it robustness \citep{nummeth}.



We use three separate terms for the parameters in the models: the true value, best-fit value, and the optimum value. The ``true" value is the initial guess supplied to the maximum-likelyhood fit, the best-fit value is the maximum likelihood best fit, fed in turn to the MCMC, and the optimum value is the value corresponding to the peak of the distribution found by MCMC.

\subsection{Error Analysis}
\label{s:errors}

\begin{equation}\label{error on d}
\sigma_d = 0.461\times d \times\sigma_\mu,
\end{equation}
with $d$ the distance and $\sigma_\mu$ the error on the distance modulus (Equation \ref{distance modulus}).

\begin{equation}\label{distance modulus}
\mu = m-M
\end{equation}

\begin{equation}\label{error on length}
\sigma_{L} = 2\tan(\frac{3\pi \times 360}{60})\times \sigma_d
\end{equation}

\begin{equation}\label{error on rho}
\sigma_\rho = \frac{2\times \sigma_{L}}{L^3}
\end{equation}
We also need to compute an error on the subtype. This is done as follows with the photometric errors $\sigma_V$ and $\sigma_J$.
\begin{equation}\label{error on type}
\sigma_{Mtype} = \sqrt[]{3.39^2\,\sigma_V^2+(-3.39)^2\,\sigma_J^2}
\end{equation}

%

\subsection{Disk+Halo Fit}
\label{s:halo and disk} 

We perform a fit with the halo-disk model. We have also ruled out pure-halo and pure-disk models, based on the unphysical or unrealistic parameter values and a highly quantized posterior (the sign of an unmixing Markov chain). 
The true parameters are still the same as in the separate fits described above. 
The optimized parameters can be found in Table \ref{tab:optimized values HD}. For the fit with all of the parameters we get the results found in Table \ref{tab:optimized values HD}. These seem reasonable apart from the value for $f_{h}$ which is unexpectedly small. Figure \ref{fig:CornerplotHD} shows the corner plot made with data from subtype 0 to 9. It shows a degeneracy between $f_h$ and $\delta{D}$. $h$--$\rho_0$, $p$--$\kappa$ and $h$--$\delta{D}$ also show some degeneracy, notably the distribution of $\rho_0$ is very peaked.

\subsubsection{Fixing Parameters}
\label{s: (disk+halo)fixed parameters} 

To improve the results of the first fit, we fix parameters of which the value is well known or of which the value is hard to constrain with the used data. The latter is the case with the scale length. Because all the lines-of-sight in our dataset are out of the plane of the Galactic disk (Figure \ref{fig:Distribution}), it is difficult to find a constraint on the scale length of the disk. Therefore we take a fixed value for the scale length at 2.6 kpc, as was found by \cite{Juric08}.  Figure \ref{fig:fixed h HD} shows the corresponding corner plot. We can see the degeneracy between $f_h$ and $\delta{D}$ and a degeneracy between $\delta{D}$ and $\rho_0$ has appeared. $\kappa$--$p$ and $f_{h}$--$\rho_0$ show some degeneracy. The distribution of $\rho_0$ is not peaked in contrast what we saw in Figure \ref{fig:CornerplotHD}. 
Adding additional constraints did not show real improvement in comparison with the fit with all parameters free and the fit with the scale length fixed. We fixed the power-law index $p$ and the central number density $\rho_0$, but there was no improvement in doing so. The results for the fit with the scale length seems to be the best. 

\subsubsection{Fit as a function of M-dwarf subtype}
\label{s:fit M-type halo+disk}

The halo-disk model with $h$ fixed at $2.6$ kpc yields the best model. With the results from this fit we find out if there is a correlation between the fit parameters and the M-dwarf subtypes. We do this for subtype $0$ to $5$ because we have too little data on the later and dimmer subtypes (Table \ref{tab:number of dwarfs}). Figure \ref{fig:subtype,z0} shows a subtype-parameter plot for the parameter $z_0$. 
There is no obvious correlation of scale-height ($z_0$) with M-dwarf subtype. This is also the case for every other parameter. In addition, when calculating the errors per subtype, we see that they are too large for the subtype-parameter plots to be reliable. This will be further discussed in section \ref{s:errors on subtype}.

\subsubsection{A check on Degenerate Parameters}

\begin{figure*}
\centering
\includegraphics[width=\linewidth]{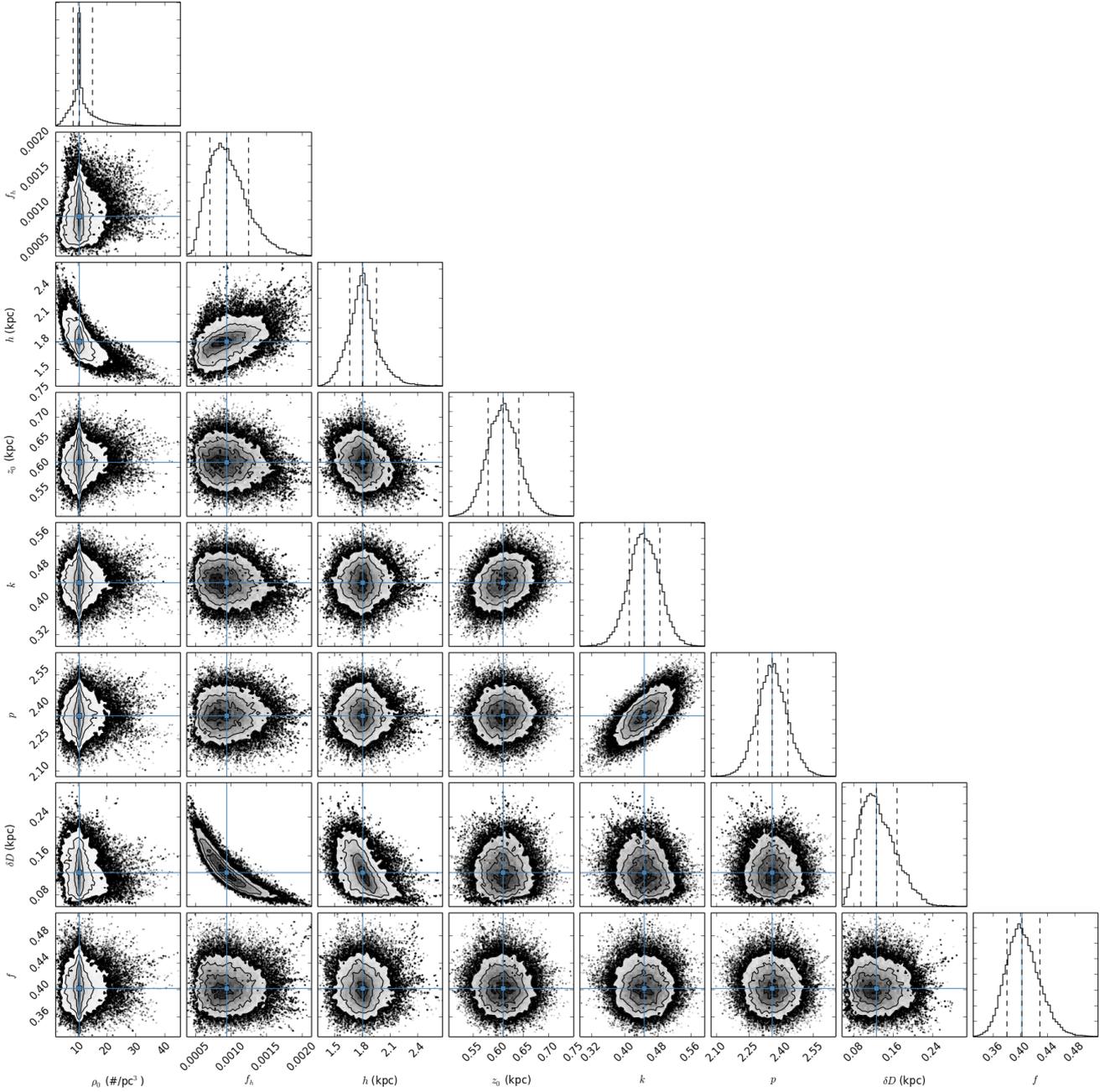}
\caption{Halo-disk model: corner plot containing subtypes M0 up to and including M9. The dotted lines give the 16th and 84th percentiles which are used for the uncertainties \protect\citep{triangle}. }
\label{fig:CornerplotHD}
\end{figure*}

\begin{figure*}
\centering
\includegraphics[width=\linewidth]{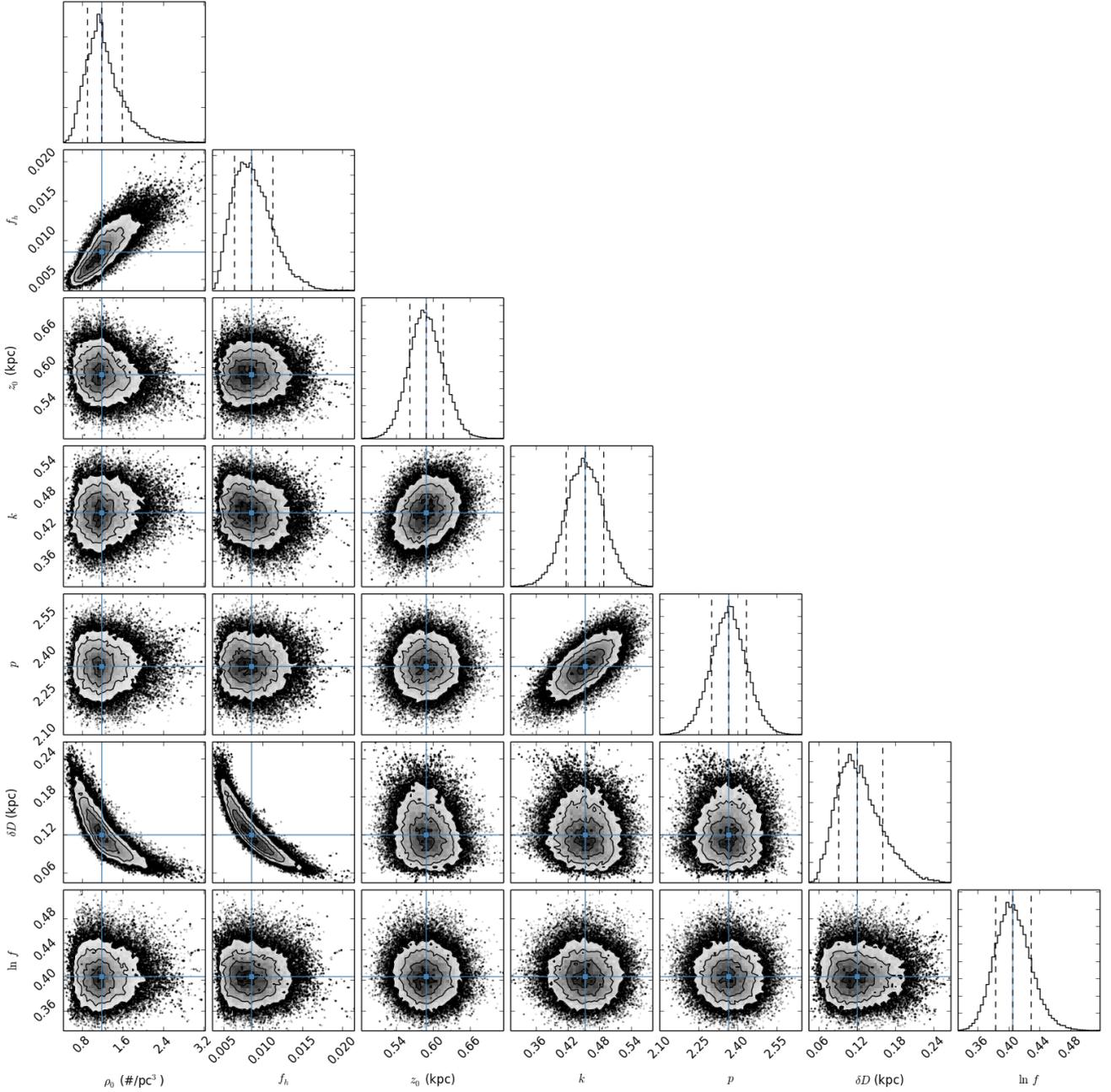}
\caption{Halo-disk model ($h$ = 2.6 kpc): corner plot made with the scale length as a fixed parameter for subtypes M0 to M9. $\rho_0$--$\delta{D}$ and $f_h$--$\delta{D}$ are degenerate. $\rho_0$--$f_h$ and $\kappa$--$p$ show some degeneracy.}
\label{fig:fixed h HD}
\end{figure*}

As a check on the effect of degenerate parameters, we change the starting ``true" parameter value of $\delta{D}$ from 1 pc into 100 pc in the fit of the halo-disk model. Table \ref{tab:global minimum} summarizes the results of the two MCMC runs. 
The biggest differences are found for $\delta{D}$, $f_{h}$ and $\rho_0$.  This can be explained by the degeneracies with $\delta{D}$ that show up in Figure \ref{fig:CornerplotHD}. The values of the total likelihoods are not the same: -2072.23 is found for $\delta{D}$ = 100 pc and -2080.02 is found for $\delta{D}$ = 1 pc but similar enough to be explained by these parameter degeneracies, which can result in multiple 'global' optima.
%

\begin{table}
\begin{tabular}{|l|c|c| c}
\hline
Parameter & Optimized value & Best fit value & Unit\\
\hline
\hline
$z_0$ 		& 0.61 kpc	 		& 0.61$\pm{0.03}$ 									& kpc\\
$h$ 			& 1.81 kpc 			& 1.79$\pm{0.14}$ 									& kpc \\
$\rho_0$ 		& 10.42 		 		&   10.52$^{+4.07}_{-1.57}$ 							& \#/$pc^3$ \\
$\kappa$ 		& 0.45 				& 0.45$\pm{0.04}$									& \\
$p$ 			& 2.36 				& 2.36$\pm{0.07}$									&  \\
$f_{h}$ 		&  9.16$\cdot 10^{-4}$ 	&  $9.08\cdot10^{-4}$ $^{+3\cdot 10^{-4}}_{-2\cdot 10^{-4}}$ 	&  \\
$\delta{D}$ 	&  0.13  	 			&  0.13$^{+0.04}_{-0.03}$								& kpc\\
$ln$f 			& 0.40 				& 0.40$^{+0.03}_{-0.02}$								& \\
\hline
\hline
\end{tabular}
\caption{\label{tab:optimized values HD} Halo-disk model: optimized values and best fit values. The $z_0$ and $\rho_0$ values are direct from the model in equation \ref{halo and disk model}. To compare these to exponential models, $z_0$ and $\rho_0$ need to decided by 2 and 4 respectively. }
\end{table}

\begin{table}
\begin{tabular}{|l|c|c| c}
\hline
Parameter & Optimized value & Best fit value & Unit\\
\hline
\hline
$z_0$ 		& 0.3  			& $0.29^{+0.02}_{-0.019}$ 			& kpc\\
$\rho_0$ 		& 0.3  		 	& $0.29^{+0.20}_{-0.13}$ 				& \#/$pc^3$ \\
$\kappa$ 		& 0.45 			& 0.45$^{+0.036}_{-0.036}$			& \\
$p$ 			& 2.37 			& 2.37$^{+0.068}_{-0.069}$			&  \\
$f_{h}$ 		&  0.0072 			& 0.0075 $^{+0.0025}_{-0.0019}$ 		& \\
$\delta{D}$ 	& 0.32 	 		& 0.31 $^{+ 0.098}_{-0.076}$ 			& kpc\\
$ln$f 			& 0.40 			& 0.40$^{+0.024}_{-0.022}$			& \\
\hline
\hline
\end{tabular}
\caption{\label{tab:optimized values HD, h=2.6} Halo-disk model ($h$ = 2.6 kpc): optimized values and best fit values. Because these are our final model values, we divided $z_0$ by 2 and $\rho_0$ by a factor 4 to facilitate comparisons with exponential models in the Literature. }
\end{table}

\begin{table}
\begin{tabular}{|l|r|r| r}
\hline
parameter & $\delta{D}$ = 100 pc & $\delta{D}$ = 1 pc & Unit\\
\hline
\hline
$2\times z_0$ 	 	& 0.61  				& 0.62 			& kpc \\
$h$ 				& 1.81  				& 2.42 			& kpc\\
$4\times \rho_0$ 	& 10.42 		 		& 39.85 			& \#/pc$^3$ \\
$\kappa$			& 0.45 				& 0.53			& \\
$p$ 				& 2.36 				& 2.57			& \\
$f_{h}$ 			&  9.16$\times 10^{-4}$ 	& 1.13$\times 10^{-2}$ & \\
$\delta{D}$ 		&  0.13  				& 9.54$\times 10^{-3}$ & kpc\\
$ln$f 			& -0.92 				& -0.85			& \\
\hline
\hline
\end{tabular}
\caption{\label{tab:global minimum} Found optimized values with initial values for $\delta{D}$ set for 1 and 100 pc for the disk+halo model with an unconstrained scale-length.}
\end{table}

\begin{table}
\begin{tabular}{|l|c|}
\hline
M-type & Number \\
\hline
\hline
0 & 33\\
1 & 31\\
2 & 19\\
3 & 16\\
4 & 49\\
5 & 39\\
6 & 3\\
7 & 7\\
8 & 7\\
9 & 5\\
10 & 1\\
\hline
\hline
\end{tabular}
\caption{\label{tab:number of dwarfs} Number of dwarfs by type in BoRG data set.}
\end{table}


\newpage

\begin{figure}
\centering
\includegraphics[width=\linewidth]{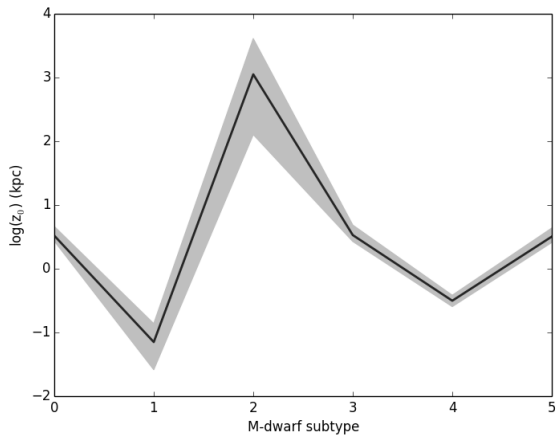}
\caption{The Halo+disk model for each M-dwarf subtype (fixed scale-length $h$ = 2.6 kpc). Shaded area indicates the $1\sigma$ uncertainty. No clear gradual relation is evident.}
\label{fig:subtype,z0}
\end{figure}

\subsection{Volume Density} 
\label{s:volume density} 

The volume density of M-dwarfs as a function of height above the plane of the Milky Way disk plots are given in Figure \ref{fig:volume density, halo+disk}. The plots are made with the best value found for the parameters. If, for example, the volume density distribution is forced into the shape of the disk-only model, the vertical distribution is too wide to represent anything that one could consider a disk (10s of kpc scale-heights). Similarly, if one considers  the distribution of M-dwarfs in the shape of the halo-only model, the distribution does not look natural. This is in contrast with Figure \ref{fig:volume density, halo+disk} where the distribution does seem reasonable (cf Figure \ref{fig:rawheights} and \ref{fig:volume density, halo+disk}). This indicates that the fit of the halo with $\kappa$ fixed and fit of the halo-disk model with h fixed are more reliable.

All of the plots also show that the later subtype brown dwarfs (M6-M9) are found near the plane ($z = 0$) of the disk, while one would expect most of them to be in the halo because of their age. This can be explained by the fact that we are not as sensitive to the dimmer M6-M9 dwarfs in the halo as we are to the earlier subtypes.

\begin{figure}
\centering
\includegraphics[width=0.5\textwidth]{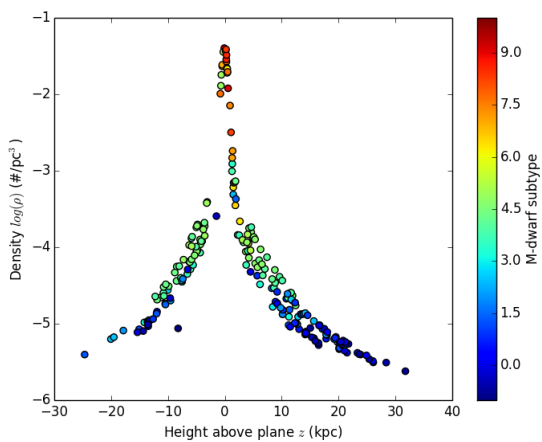}
\caption{Halo-disk model ($h$ = 2.6): volume density of M-dwarfs as a function as height above the plane of the Milky Way disk. 
Compare this model distribution to photometric positions in Figure \ref{fig:rawheights}.}
\label{fig:volume density, halo+disk}
\end{figure}

\subsection{\label{s:number and mass}Total number and mass of M-dwarfs}


The total number of M-dwarfs in the disk can be determined by integrating Equation \ref{model} over the cylindrical coordinates $R$, $z$ and $\theta$. The limits chosen are commonly used for the limits of our Galaxy.
\begin{equation}\label{total number disk}
N = \rho_0 \int\limits_{0}^{2\pi} \int\limits_{-10z_0}^{10z_0} \int\limits_{0}^{10h} e^{-\frac{R}{h}}\sech^2\left(\frac{z}{z_0}\right)R\mathrm{d}R\mathrm{d}z\mathrm{d}\theta
\end{equation}
{The total mass of M-dwarfs in the disk is computed with Equation} \ref{total mass disk}. 
{For $m_0$ we use $70 M_J$ and $600 M_J$ to calculate a lower and upper limit for the mass} \citep{Reid04, Kaltenegger09}. 
\begin{equation}\label{total mass disk}
M = 2 m_0\rho_0 \int\limits_{0}^{2\pi} \int\limits_{0}^{10z_0} \int\limits_{0}^{10h} e^{-\frac{R}{h}}\sech^2\left(\frac{z}{z_0}\right)R\mathrm{d}R\mathrm{d}z\mathrm{d}\theta
\end{equation}
The total number of M-dwarfs in the halo can be determined by integrating Equation \ref{halo model}. 
The integral becomes:
\begin{equation}\label{total number halo}
N = 2 \int\limits_{0}^{2\pi} \int\limits_{0}^{10z_0} \int\limits_{0}^{10h}\rho_{\odot}\,f_{h}\,{\left(\frac{R^2+(z/\kappa)^2}{R_{\odot}^2+(z_{\odot}/\kappa)^2}\right)}^{-p/2}R\mathrm{d}R\mathrm{d}z\mathrm{d}\theta
\end{equation}
The total mass of M-dwarfs in the halo is computed with Equation \ref{total mass halo}.
\begin{equation}\label{total mass halo}
M = 2 m_0\int\limits_{0}^{2\pi} \int\limits_{0}^{10z_0} \int\limits_{0}^{10h}\rho_{\odot}\,f_{h}\, \left( {\frac{R^2+(z/\kappa)^2}{R_{\odot}^2+(z_{\odot}/\kappa)^2}} \right)^{-p/2}R\mathrm{d}R\mathrm{d}z\mathrm{d}\theta
\end{equation}
{The resulting inferred number of M-dwarfs in the Milky Way disk and their lower and upper mass estimates are summarized in Table }\ref{tab:number and mass halo}. {A total of $\sim10^9 M_\odot$ of the mass of the Milky Way disk is in $\sim58\times10^9$ M-type brown dwarf members. }

\begin{table}
\begin{tabular}{|l|c|c|c|}
\hline
 &  & + & - \\
\hline
\hline 
Number 						& $58.2 \cdot 10^{9}$ 	& 9.81$\cdot10^{9}$ & 6.70$\cdot10^{9}$\\
{ Lower limit mass ($M_{\odot}$) }	& 4.26$\cdot10^{9}$ 	& 0.69$\cdot10^{9}$ & 0.47$\cdot10^{8}$\\
{Upper limit mass ($M_{\odot}$) }	& 36.52$\cdot10^{9}$ 	& 5.89$\cdot10^{9}$ & 4.02$\cdot10^{9}$\\
\hline
\hline
\end{tabular}
\caption{\label{tab:number and mass halo} Total number and mass of M-dwarfs in the halo and disk of the Milky Way.}
\end{table}

\begin{figure}
\includegraphics[width=0.9\linewidth]{figures/fraction.pdf}
\captionof{figure}{The mass fraction in the stellar halo as a function of the total stellar mass. The red line is the predicted median relation between the accreted mass fraction and the total stellar mass from \protect\cite{Cooper13}. The green and orange line indicate the respectively the 1$\sigma$ and 2$\sigma$ limits. Also displayed are the values found for the Milky Way \protect\citep{Courteau11}, M31 \protect\citep{Ibata}, M81 \protect\citep{Barker09}, M253 \protect\citep{Bailin11}, M101 \protect\citep{van-Dokkum14}, N891, N5236 and N4565 \protect\citep{GHOSTS}.} 
\label{fig:fraction}
\end{figure}

\subsection{Halo Fraction}
\label{s:fraction halo}

We can find the fraction of halo M-dwarfs and compare it with the numerical simulations from \cite{Cooper13}. They give the relation between stellar mass accreted early through galaxy mergers and the total stellar mass. 

\setlength{\parindent}{6mm}The fraction of halo stars we found is 7$^{+5}_{-4}$\%, higher than the 2\% fraction found by \cite{Courteau11}. In Figure \ref{fig:fraction} we display the found value for the fraction and the total stellar mass of the Milky Way of \cite{Courteau11} with the model of \cite{Cooper13}. We see that our value found for the halo fraction of the Milky Way lies within the margins of the model. We note that our disk+halo model does not include the bulge (or a thick disk component) and so our fraction of 7\% is an upper limit of the halo fraction of stars. 
Assuming a 15\% bulge contribution to the bulge+disk galaxy total, this halo fraction would be 6\%. We note that the 7\% value is uncertain and it is still in agreement within $2\sigma$ with the  \cite{Courteau11} value.

\section{Discussion}
\label{s:discussion}

\subsection{Errors on subtype}
\label{s:errors on subtype}

The errors on the subtypes of brown dwarfs are calculated with Equation \ref{error on type}. To find the average error on one subtype, the photometric errors in the color calculation are quadratically added 
for all the stars of that subtype. We perform this check to verify if a fit per subtype (Section \ref{s:fit M-type halo+disk}) is feasible. The results of these type error can be found in Table \ref{tab: error on type}.
The found average type errors are quite large, especially the ones on subtype M0, M2 and M9. Any relation found with the fit per subtype is therefore doubtful. A brown dwarf subtyped as M2 could well be any other subtype with these errors. 

There are a few large errors on the $F606W$ filter magnitude which cause these large values. The magnitude errors could also be due to the fact that used data was taken with two different filters ($F606W$ and $F600LP$ for the BoRG and HIPPIES survey respectively). The photometry in \cite{Holwerda14} was done with $F606W$, but because part of the fields were not observed with this particular filter, a correction had to be made. Three fields were observed with both of the filters in question. They found the colour difference and corrected the magnitudes for $F606W$ and increased the respective errors. 

However, the close relation between optical-infrared color and subtype observed by \cite{Holwerda14} in CANDELS data raises the possibility that with higher-fidelity photometry and/or multi-band photometry, brown dwarf photometric typing and sub-typing may be feasible in the future. 


\subsection{Comparison}
\label{s:comparison}

{When we compare the found values with the best working model (halo-disk model) with values from earlier research they are in excellent agreement (Table} \ref{tab:comparison}).{ Especially the scale-height found here compares well with those for other types of brown dwarfs in other surveys (e.g.,} \cite{Pirzkal09} or \cite{Ryan11}) {with slightly improved errors thanks to the multiple lines-of-sight, the MCMC approach and the inclusion of the second, halo component.}

We constrain the halo parameters better than the 2MASS survey model, thanks to the depth of the data and the unambiguous identification of stellar objects and M-dwarfs in HST data. We can attribute this to the trade between depth and survey area resulting in a relatively large survey volume which is more optimized for the halo.

\begin{table}
\begin{tabular}{|l|r|}
\hline
M-type & subtype error \\
\hline
\hline
0 & 6.87\\
1 & 0.35\\
2 &  17.45\\
3 & 1.18\\
4 &  2.87\\
5 & 11.43\\
\hline
\hline
\end{tabular}
\caption{\label{tab: error on type}Error on subtypes.}
\end{table}

\begin{table*}
\resizebox{1.\textwidth}{!}{
\begin{tabular}{|l|c|c|c|c|c|c| c}
\hline
Parameter 					& Our value 			&  \cite{Juric08} 		& \cite{Pirzkal09} 		& \cite{Ryan11} 		& \cite{Zheng01} 	& \cite{Chang}		& Unit \\
\hline
\hline
Scale height ($z_0$)			& 0.29 $\pm{0.02}$   	& - 					& - 					& 0.3$\pm{0.056}$ 	 	& - 				& - 				& kpc\\
Scale height ($z_0$) (thin disk)  	& - 					& 0.3$\pm{0.06}$  	 	& 0.3$\pm{0.07}$ 	 	& - 					& - 				& 0.36$\pm{0.01}$ 	& kpc \\
Scale height ($z_0$) (thick disk)  	& - 					& 0.9$\pm{0.18}$   	 	& - 					& - 					& - 				& 1.02$\pm{0.03}$ 	& kpc\\
Scale length (h) 				& - 					& - 					& - 					&  - 					& 2.75$\pm{0.41}$ 	& -				& kpc\\
Scale length (h) (thin disk)   		& - 					& 2.6$\pm{0.52}$ 	 	& - 					&  - 					& - 				& 3.7$\pm{1.0}$ 	& kpc\\
Scale length (h) (thick disk)   		& - 					& 3.6$\pm{0.72}$ 	 	& - 					&  - 					& - 				& 5.0$\pm{1.0}$ 	& kpc\\
Central number density ($\rho_0$) 	&  0.29 $^{+0.20}_{-0.13}$ & - 					& - 					& - 					& - 				& -				&  \#/pc$^3$ \\
Flattening parameter ($\kappa$) 	& 0.45 $\pm{0.04}$ 		& 0.64$\pm{0.13}$ 		& - 					& - 					& - 				& 0.55$\pm{0.15}$	& \\
Power-law index($p$) 			&  2.4 $\pm{0.07}$ 		& 2.8$\pm{0.56}$ 		& - 					& - 					& - 				& 2.6$\pm{0.6}$	& \\
Fraction ($f_{h}$) 				& 0.0075$^{+0.0025}_{-0.0019}$ & 0.005$\pm{0.001}$ & - 				& - 					& - 				& 0.002$\pm{0.001}$ & \\
\hline
\hline
\end{tabular}
}
\caption{\label{tab:comparison} Our best values compared to earlier found values by \protect\cite{Juric08}, \protect\cite{Pirzkal09}, \protect\cite{Ryan11}, \protect\cite{Zheng01} and \protect\cite{Chang}. We note that in order to compare we report the $\rho_0/4$ and $z_0/2$ values from the MCMC fit.}
\end{table*}

\section{Conclusions} 
\label{s:conclusions}


From our MCMC model of the number of M-dwarf stars found the in BoRG survey, we conclude:
\begin{enumerate}
\item The disk+halo model works best, a single component model results in unphysical results.
\item We found the following values for the parameters:
	\begin{enumerate}
	\item The scale height: $z_0$ = 0.29$\pm{0.02}$ kpc
    \item The central number density: $\rho_0$ = 0.29$^{+0.20}_{-0.13}$ \#/pc$^3$
    \item The power law index: $p$ = 2.4$\pm{0.07}$
    \item The flattening parameter: $\kappa$ = 0.45$\pm{0.04}$
    \item The local halo fraction: $f_{h}$ = 0.0075$^{+0.0025}_{-0.0019}$
	\item The bin-width: $\delta{D}$ = 0.31$^{+0.09}_{-0.08}$ kpc
\end{enumerate}
\item We found that $f_{h}$--$\delta{D}$ and $\kappa$--$p$ are degenerate. We could not find any correlation between subtype and parameter.
\item The total number of M-dwarfs in the halo and disk is: $58.2$ $^{+9.3}_{-6.2}\times10^{9}$. The upper limit of the halo fraction is 7$^{+5}_{-4}$\%.
\item The upper and lower limit of the total mass of M-dwarfs in the halo and disk are respectively: $M_{upper}$ = $1.99$ $^{+0.73}_{-0.50}\times10^{9}$ $M_{\odot}$ and $M_{lower}$ = $0.32$ $^{+0.12}_{-0.081} \times10^{9}$ $M_{\odot}$.
\end{enumerate}

\section{Future Work}
\label{s:future}

The counts in random HST fields of the lowest-mass stars and brown dwarfs (to model their distribution and number in the Milky Way) will serve in future years for two important new astronomical space missions, EUCLID and JWST.
 EUCLID will map most of the sky to a similar depth, spatial resolution, and filters as the BoRG survey.  
 With improved brown dwarfs statistics, the current model will become an more accurate measure for the shape and structure of the Milky Way disk and halo. The EUCLID mission will be able to detect nearly all streams and satellite galaxies of the Milky Way: all halo substructure can be detected using these objects as the tracer \citep{EUCLID}. Given their ubiquity, stellar overdensities stand out in greater contrast \citep{Holwerda14}.
 Once a full tally has been made, the implications for the Galaxy-wide and halo IMF can be explored. 
 JWST is a precision near-infrared observatory. To accurately map spectroscopic instruments such as NIRspec, onboard JWST, onto distant targets, a multitude of NIR-bright reference points will be needed (Holwerda et al. {\em in prep}). We developed in part this model of the Milk Way Disk and Halo, to aid in the predicted numbers of guide stars in JWST imaging.

\section*{Acknowledgements}

The authors would like to thank the referee for his or her outstanding and helpful referee rapport which improved the manuscript substantially.
%
%
This research made use of Astropy, a community-developed core Python package for Astronomy \citep{Astropy-Collaboration13a}. This research made use of matplotlib, a Python library for publication quality graphics \citep{Hunter07}. PyRAF is a product of the Space Telescope Science Institute, which is operated by AURA for NASA. This research made use of SciPy \citep{scipy}. 
This research has made use of the NASA/IPAC Extragalactic Database (NED) which is operated by the Jet Propulsion Laboratory, California Institute of Technology, under contract with the National Aeronautics and Space Administration. 
This research has made use of NASA's Astrophysics Data System.

\bibliographystyle{mnras}
\bibliography{Bibliography}

\begin{thebibliography}{}
\makeatletter
\relax
\def\mn@urlcharsother{\let\do\@makeother \do\$\do\&\do\#\do\^\do\_\do\%\do\~}
\def\mn@doi{\begingroup\mn@urlcharsother \@ifnextchar [ {\mn@doi@}
  {\mn@doi@[]}}
\def\mn@doi@[#1]#2{\def\@tempa{#1}\ifx\@tempa\@empty \href
  {http://dx.doi.org/#2} {doi:#2}\else \href {http://dx.doi.org/#2} {#1}\fi
  \endgroup}
\def\mn@eprint#1#2{\mn@eprint@#1:#2::\@nil}
\def\mn@eprint@arXiv#1{\href {http://arxiv.org/abs/#1} {{\tt arXiv:#1}}}
\def\mn@eprint@dblp#1{\href {http://dblp.uni-trier.de/rec/bibtex/#1.xml}
  {dblp:#1}}
\def\mn@eprint@#1:#2:#3:#4\@nil{\def\@tempa {#1}\def\@tempb {#2}\def\@tempc
  {#3}\ifx \@tempc \@empty \let \@tempc \@tempb \let \@tempb \@tempa \fi \ifx
  \@tempb \@empty \def\@tempb {arXiv}\fi \@ifundefined
  {mn@eprint@\@tempb}{\@tempb:\@tempc}{\expandafter \expandafter \csname
  mn@eprint@\@tempb\endcsname \expandafter{\@tempc}}}

\bibitem[\protect\citeauthoryear{{Aberasturi}, {Burgasser}, {Mora}, {Solano},
  {Mart{\'{\i}}n}, {Reid}  \& {Looper}}{{Aberasturi}
  et~al.}{2014}]{Aberasturi14}
{Aberasturi} M.,  {Burgasser} A.~J.,  {Mora} A.,  {Solano} E.,  {Mart{\'{\i}}n}
  E.~L.,  {Reid} I.~N.,   {Looper} D.,  2014, \mn@doi [\aj]
  {10.1088/0004-6256/148/6/129}, \href
  {http://adsabs.harvard.edu/abs/2014AJ....148..129A} {148, 129}

\bibitem[\protect\citeauthoryear{{Astropy Collaboration} et~al.,}{{Astropy
  Collaboration} et~al.}{2013a}]{Astropy}
{Astropy Collaboration} et~al., 2013a, \mn@doi [\aap]
  {10.1051/0004-6361/201322068}, \href
  {http://adsabs.harvard.edu/abs/2013A%26A...558A..33A} {558, A33}

\bibitem[\protect\citeauthoryear{{Astropy Collaboration} et~al.,}{{Astropy
  Collaboration} et~al.}{2013b}]{Astropy-Collaboration13a}
{Astropy Collaboration} et~al., 2013b, \mn@doi [\aap]
  {10.1051/0004-6361/201322068}, \href
  {http://adsabs.harvard.edu/abs/2013A%26A...558A..33A} {558, A33}

\bibitem[\protect\citeauthoryear{{Bailin}, {Bell}, {Chappell}, {Radburn-Smith}
  \& {de Jong}}{{Bailin} et~al.}{2011}]{Bailin11}
{Bailin} J.,  {Bell} E.~F.,  {Chappell} S.~N.,  {Radburn-Smith} D.~J.,   {de
  Jong} R.~S.,  2011, \mn@doi [\apj] {10.1088/0004-637X/736/1/24}, \href
  {http://adsabs.harvard.edu/abs/2011ApJ...736...24B} {736, 24}

\bibitem[\protect\citeauthoryear{{Barker}, {Ferguson}, {Irwin}, {Arimoto}  \&
  {Jablonka}}{{Barker} et~al.}{2009}]{Barker09}
{Barker} M.~K.,  {Ferguson} A.~M.~N.,  {Irwin} M.,  {Arimoto} N.,   {Jablonka}
  P.,  2009, \mn@doi [\aj] {10.1088/0004-6256/138/5/1469}, \href
  {http://adsabs.harvard.edu/abs/2009AJ....138.1469B} {138, 1469}

\bibitem[\protect\citeauthoryear{{Beckwith} et~al.,}{{Beckwith}
  et~al.}{2006}]{Beckwith06}
{Beckwith} S.~V.~W.,  et~al., 2006, \mn@doi [\aj] {10.1086/507302}, \href
  {http://adsabs.harvard.edu/abs/2006AJ....132.1729B} {132, 1729}

\bibitem[\protect\citeauthoryear{{Belokurov} et~al.,}{{Belokurov}
  et~al.}{2006}]{Belokurov06}
{Belokurov} V.,  et~al., 2006, \mn@doi [\apjl] {10.1086/504797}, \href
  {http://adsabs.harvard.edu/abs/2006ApJ...642L.137B} {642, L137}

\bibitem[\protect\citeauthoryear{{Bensby}, {Alves-Brito}, {Oey}, {Yong}  \&
  {Mel{\'e}ndez}}{{Bensby} et~al.}{2011}]{Bensby11}
{Bensby} T.,  {Alves-Brito} A.,  {Oey} M.~S.,  {Yong} D.,   {Mel{\'e}ndez} J.,
  2011, \mn@doi [\apjl] {10.1088/2041-8205/735/2/L46}, \href
  {http://adsabs.harvard.edu/abs/2011ApJ...735L..46B} {735, L46}

\bibitem[\protect\citeauthoryear{Binney \& Tremaine}{Binney \&
  Tremaine}{2008}]{BT1}
Binney J.,  Tremaine S.,  2008, {Galactic Dynamics}, second edn.
Princeton Series in Astrophysics, Princeton University Press

\bibitem[\protect\citeauthoryear{{Bochanski}, {Willman}, {West}, {Strader}  \&
  {Chomiuk}}{{Bochanski} et~al.}{2014}]{Bochanski14}
{Bochanski} J.~J.,  {Willman} B.,  {West} A.~A.,  {Strader} J.,   {Chomiuk} L.,
   2014, \mn@doi [\aj] {10.1088/0004-6256/147/4/76}, \href
  {http://adsabs.harvard.edu/abs/2014AJ....147...76B} {147, 76}

\bibitem[\protect\citeauthoryear{{Bovy}, {Rix}, {Liu}, {Hogg}, {Beers}  \&
  {Lee}}{{Bovy} et~al.}{2012}]{Bovy12b}
{Bovy} J.,  {Rix} H.-W.,  {Liu} C.,  {Hogg} D.~W.,  {Beers} T.~C.,   {Lee}
  Y.~S.,  2012, \mn@doi [\apj] {10.1088/0004-637X/753/2/148}, \href
  {http://adsabs.harvard.edu/abs/2012ApJ...753..148B} {753, 148}

\bibitem[\protect\citeauthoryear{{Bradley} et~al.,}{{Bradley}
  et~al.}{2012}]{Bradley12}
{Bradley} L.~D.,  et~al., 2012, \mn@doi [\apj] {10.1088/0004-637X/760/2/108},
  \href {http://adsabs.harvard.edu/abs/2012ApJ...760..108B} {760, 108}

\bibitem[\protect\citeauthoryear{{Chang}, {Ko}  \& {Peng}}{{Chang}
  et~al.}{2011}]{Chang}
{Chang} C.-K.,  {Ko} C.-M.,   {Peng} T.-H.,  2011, \mn@doi [\apj]
  {10.1088/0004-637X/740/1/34}, \href
  {http://adsabs.harvard.edu/abs/2011ApJ...740...34C} {740, 34}

\bibitem[\protect\citeauthoryear{{Chen} et~al.,}{{Chen} et~al.}{2001}]{Chen01}
{Chen} B.,  et~al., 2001, \mn@doi [\apj] {10.1086/320647}, \href
  {http://adsabs.harvard.edu/abs/2001ApJ...553..184C} {553, 184}

\bibitem[\protect\citeauthoryear{{Cheng} et~al.,}{{Cheng}
  et~al.}{2012}]{Cheng12}
{Cheng} J.~Y.,  et~al., 2012, preprint, \href
  {http://adsabs.harvard.edu/abs/2012arXiv1204.5179C} {} (\mn@eprint {arXiv}
  {1204.5179})

\bibitem[\protect\citeauthoryear{{Cooper}, {D'Souza}, {Kauffmann}, {Wang},
  {Boylan-Kolchin}, {Guo}, {Frenk}  \& {White}}{{Cooper}
  et~al.}{2013}]{Cooper13}
{Cooper} A.~P.,  {D'Souza} R.,  {Kauffmann} G.,  {Wang} J.,  {Boylan-Kolchin}
  M.,  {Guo} Q.,  {Frenk} C.~S.,   {White} S.~D.~M.,  2013, \mn@doi [\mnras]
  {10.1093/mnras/stt1245}, \href
  {http://adsabs.harvard.edu/abs/2013MNRAS.434.3348C} {434, 3348}

\bibitem[\protect\citeauthoryear{{Courteau}, {Widrow}, {McDonald},
  {Guhathakurta}, {Gilbert}, {Zhu}, {Beaton}  \& {Majewski}}{{Courteau}
  et~al.}{2011}]{Courteau11}
{Courteau} S.,  {Widrow} L.~M.,  {McDonald} M.,  {Guhathakurta} P.,  {Gilbert}
  K.~M.,  {Zhu} Y.,  {Beaton} R.~L.,   {Majewski} S.~R.,  2011, \mn@doi [\apj]
  {10.1088/0004-637X/739/1/20}, \href
  {http://adsabs.harvard.edu/abs/2011ApJ...739...20C} {739, 20}

\bibitem[\protect\citeauthoryear{{Ford}}{{Ford}}{2006}]{Ford}
{Ford} E.~B.,  2006, \mn@doi [\apj] {10.1086/500802}, \href
  {http://adsabs.harvard.edu/abs/2006ApJ...642..505F} {642, 505}

\bibitem[\protect\citeauthoryear{{Foreman-Mackey}, {Hogg}, {Lang}  \&
  {Goodman}}{{Foreman-Mackey} et~al.}{2013}]{MCMC}
{Foreman-Mackey} D.,  {Hogg} D.~W.,  {Lang} D.,   {Goodman} J.,  2013, \mn@doi
  [\pasp] {10.1086/670067}, \href
  {http://adsabs.harvard.edu/abs/2013PASP..125..306F} {125, 306}

\bibitem[\protect\citeauthoryear{Foreman-Mackey, Price-Whelan, Ryan, Emily,
  Smith, Barbary, Hogg  \& Brewer}{Foreman-Mackey et~al.}{2014}]{triangle}
Foreman-Mackey D.,  Price-Whelan A.,  Ryan G.,  Emily Smith M.,  Barbary K.,
  Hogg D.~W.,   Brewer B.~J.,  2014, {triangle.py v0.1.1},
  \mn@doi{10.5281/zenodo.11020}

\bibitem[\protect\citeauthoryear{{Giavalisco} et~al.,}{{Giavalisco}
  et~al.}{2004}]{goods}
{Giavalisco} M.,  et~al., 2004, \mn@doi [\apjl] {10.1086/379232}, \href
  {http://adsabs.harvard.edu/cgi-bin/nph-bib_query?bibcode=2004ApJ...600L..93G%
&db_key=AST} {600, L93}

\bibitem[\protect\citeauthoryear{{Gilmore} \& {Reid}}{{Gilmore} \&
  {Reid}}{1983}]{Gilmore83}
{Gilmore} G.,  {Reid} N.,  1983, \mnras, \href
  {http://adsabs.harvard.edu/abs/1983MNRAS.202.1025G} {202, 1025}

\bibitem[\protect\citeauthoryear{{Gregory}}{{Gregory}}{2005}]{Gregory05}
{Gregory} P.~C.,  2005, \mn@doi [\apj] {10.1086/432594}, \href
  {http://adsabs.harvard.edu/abs/2005ApJ...631.1198G} {631, 1198}

\bibitem[\protect\citeauthoryear{{Haas}}{{Haas}}{2010}]{Haas}
{Haas} M.~R.,  2010, PhD thesis, Ph.~D.~thesis, University of Leiden (2010)

\bibitem[\protect\citeauthoryear{{Hawley} et~al.,}{{Hawley}
  et~al.}{2002}]{Hawley02}
{Hawley} S.~L.,  et~al., 2002, \mn@doi [\aj] {10.1086/340697}, \href
  {http://adsabs.harvard.edu/abs/2002AJ....123.3409H} {123, 3409}

\bibitem[\protect\citeauthoryear{{Helmi}}{{Helmi}}{2008}]{Helmi}
{Helmi} A.,  2008, \mn@doi [\aapr] {10.1007/s00159-008-0009-6}, \href
  {http://adsabs.harvard.edu/abs/2008A%26ARv..15..145H} {15, 145}

\bibitem[\protect\citeauthoryear{{Holwerda} et~al.,}{{Holwerda}
  et~al.}{2014}]{Holwerda14}
{Holwerda} B.~W.,  et~al., 2014, \mn@doi [\apj] {10.1088/0004-637X/788/1/77},
  \href {http://adsabs.harvard.edu/abs/2014ApJ...788...77H} {788, 77}

\bibitem[\protect\citeauthoryear{{Hou}, {Goodman}, {Hogg}, {Weare}  \&
  {Schwab}}{{Hou} et~al.}{2012}]{Hou12}
{Hou} F.,  {Goodman} J.,  {Hogg} D.~W.,  {Weare} J.,   {Schwab} C.,  2012,
  \mn@doi [\apj] {10.1088/0004-637X/745/2/198}, \href
  {http://adsabs.harvard.edu/abs/2012ApJ...745..198H} {745, 198}

\bibitem[\protect\citeauthoryear{Hunter}{Hunter}{2007}]{Hunter07}
Hunter J.~D.,  2007, Computing In Science \& Engineering, 9, 90

\bibitem[\protect\citeauthoryear{{Ibata} et~al.,}{{Ibata} et~al.}{2014}]{Ibata}
{Ibata} R.~A.,  et~al., 2014, \mn@doi [\apj] {10.1088/0004-637X/780/2/128},
  \href {http://adsabs.harvard.edu/abs/2014ApJ...780..128I} {780, 128}

\bibitem[\protect\citeauthoryear{{Janson} et~al.,}{{Janson}
  et~al.}{2012}]{Janson12}
{Janson} M.,  et~al., 2012, \mn@doi [\apj] {10.1088/0004-637X/754/1/44}, \href
  {http://adsabs.harvard.edu/abs/2012ApJ...754...44J} {754, 44}

\bibitem[\protect\citeauthoryear{{Jones}, {Oliphant}, {Peterson}  \&
  Others}{{Jones} et~al.}{2001}]{scipy}
{Jones} E.,  {Oliphant} T.,  {Peterson} P.,   Others 2001, {SciPy}: Open source
  scientific tools for Python, \url {http://www.scipy.org/}

\bibitem[\protect\citeauthoryear{{Juri{\'c}} et~al.,}{{Juri{\'c}}
  et~al.}{2008}]{Juric08}
{Juri{\'c}} M.,  et~al., 2008, \mn@doi [\apj] {10.1086/523619}, \href
  {http://adsabs.harvard.edu/abs/2008ApJ...673..864J} {673, 864}

\bibitem[\protect\citeauthoryear{{Kaltenegger} \& {Traub}}{{Kaltenegger} \&
  {Traub}}{2009}]{Kaltenegger09}
{Kaltenegger} L.,  {Traub} W.~A.,  2009, \mn@doi [\apj]
  {10.1088/0004-637X/698/1/519}, \href
  {http://adsabs.harvard.edu/abs/2009ApJ...698..519K} {698, 519}

\bibitem[\protect\citeauthoryear{{Kapteyn}}{{Kapteyn}}{1922}]{Kapteyn22}
{Kapteyn} J.~C.,  1922, \mn@doi [\apj] {10.1086/142670}, \href
  {http://adsabs.harvard.edu/abs/1922ApJ....55..302K} {55, 302}

\bibitem[\protect\citeauthoryear{Kiusalaas}{Kiusalaas}{2013}]{nummeth}
Kiusalaas J.,  2013, {Numerical Methods in Engineering with Python 3}, third
  edn.
Cambrigde University Press

\bibitem[\protect\citeauthoryear{{Laureijs} et~al.,}{{Laureijs}
  et~al.}{2011}]{EUCLID}
{Laureijs} R.,  et~al., 2011, preprint, \href
  {http://adsabs.harvard.edu/abs/2011arXiv1110.3193L} {} (\mn@eprint {arXiv}
  {1110.3193})

\bibitem[\protect\citeauthoryear{LeBlanc}{LeBlanc}{2010}]{LeBlanc}
LeBlanc F.,  2010, {An introduction to stellar astrophysics}, first edn.
Wiley

\bibitem[\protect\citeauthoryear{{Majewski}, {Skrutskie}, {Weinberg}  \&
  {Ostheimer}}{{Majewski} et~al.}{2003}]{Majewski03}
{Majewski} S.~R.,  {Skrutskie} M.~F.,  {Weinberg} M.~D.,   {Ostheimer} J.~C.,
  2003, \mn@doi [\apj] {10.1086/379504}, \href
  {http://adsabs.harvard.edu/abs/2003ApJ...599.1082M} {599, 1082}

\bibitem[\protect\citeauthoryear{{Oort}}{{Oort}}{1938}]{Oort38}
{Oort} J.~H.,  1938, \bain, \href
  {http://adsabs.harvard.edu/abs/1938BAN.....8..233O} {8, 233}

\bibitem[\protect\citeauthoryear{{Pirzkal} et~al.,}{{Pirzkal}
  et~al.}{2005}]{Pirzkal05}
{Pirzkal} N.,  et~al., 2005, \mn@doi [\apj] {10.1086/427896}, \href
  {http://adsabs.harvard.edu/abs/2005ApJ...622..319P} {622, 319}

\bibitem[\protect\citeauthoryear{{Pirzkal} et~al.,}{{Pirzkal}
  et~al.}{2009}]{Pirzkal09}
{Pirzkal} N.,  et~al., 2009, \mn@doi [\apj] {10.1088/0004-637X/695/2/1591},
  \href {http://adsabs.harvard.edu/abs/2009ApJ...695.1591P} {695, 1591}

\bibitem[\protect\citeauthoryear{{Radburn-Smith} et~al.,}{{Radburn-Smith}
  et~al.}{2011}]{GHOSTS}
{Radburn-Smith} D.~J.,  et~al., 2011, \mn@doi [\apjs]
  {10.1088/0067-0049/195/2/18}, \href
  {http://adsabs.harvard.edu/abs/2011ApJS..195...18R} {195, 18}

\bibitem[\protect\citeauthoryear{{Reid} et~al.,}{{Reid} et~al.}{2004}]{Reid04}
{Reid} I.~N.,  et~al., 2004, \mn@doi [\aj] {10.1086/421374}, \href
  {http://adsabs.harvard.edu/abs/2004AJ....128..463R} {128, 463}

\bibitem[\protect\citeauthoryear{{Reid}, {Cruz}, {Kirkpatrick}, {Allen},
  {Mungall}, {Liebert}, {Lowrance}  \& {Sweet}}{{Reid} et~al.}{2008}]{Reid08}
{Reid} I.~N.,  {Cruz} K.~L.,  {Kirkpatrick} J.~D.,  {Allen} P.~R.,  {Mungall}
  F.,  {Liebert} J.,  {Lowrance} P.,   {Sweet} A.,  2008, \mn@doi [\aj]
  {10.1088/0004-6256/136/3/1290}, \href
  {http://adsabs.harvard.edu/abs/2008AJ....136.1290R} {136, 1290}

\bibitem[\protect\citeauthoryear{{Rhodes}}{{Rhodes}}{2011}]{pyephem}
{Rhodes} B.~C.,  2011, {PyEphem: Astronomical Ephemeris for Python} (\mn@eprint
  {ascl} {Astrophysics Source Code Library, ascl:1112.014})

\bibitem[\protect\citeauthoryear{{Ryan}, {Hathi}, {Cohen}  \&
  {Windhorst}}{{Ryan} et~al.}{2005}]{Ryan05}
{Ryan} Jr. R.~E.,  {Hathi} N.~P.,  {Cohen} S.~H.,   {Windhorst} R.~A.,  2005,
  \mn@doi [\apjl] {10.1086/497368}, \href
  {http://adsabs.harvard.edu/abs/2005ApJ...631L.159R} {631, L159}

\bibitem[\protect\citeauthoryear{{Ryan} et~al.,}{{Ryan} et~al.}{2011}]{Ryan11}
{Ryan} R.~E.,  et~al., 2011, \mn@doi [\apj] {10.1088/0004-637X/739/2/83}, \href
  {http://adsabs.harvard.edu/abs/2011ApJ...739...83R} {739, 83}

\bibitem[\protect\citeauthoryear{{Scalo}}{{Scalo}}{1986}]{Scalo}
{Scalo} J.~M.,  1986, \fcp, \href
  {http://adsabs.harvard.edu/abs/1986FCPh...11....1S} {11, 1}

\bibitem[\protect\citeauthoryear{{Schmidt} et~al.,}{{Schmidt}
  et~al.}{2014}]{Schmidt14}
{Schmidt} K.~B.,  et~al., 2014, \mn@doi [\apj] {10.1088/0004-637X/786/1/57},
  \href {http://adsabs.harvard.edu/abs/2014ApJ...786...57S} {786, 57}

\bibitem[\protect\citeauthoryear{{Seares}, {van Rhijn}, {Joyner}  \&
  {Richmond}}{{Seares} et~al.}{1925}]{Seares25}
{Seares} F.~H.,  {van Rhijn} P.~J.,  {Joyner} M.~C.,   {Richmond} M.~L.,  1925,
  \mn@doi [\apj] {10.1086/142940}, \href
  {http://adsabs.harvard.edu/abs/1925ApJ....62..320S} {62, 320}

\bibitem[\protect\citeauthoryear{{Stanway}, {Bremer}, {Lehnert}  \&
  {Eldridge}}{{Stanway} et~al.}{2008}]{Stanway08}
{Stanway} E.~R.,  {Bremer} M.~N.,  {Lehnert} M.~D.,   {Eldridge} J.~J.,  2008,
  \mn@doi [\mnras] {10.1111/j.1365-2966.2007.12711.x}, \href
  {http://adsabs.harvard.edu/abs/2008MNRAS.384..348S} {384, 348}

\bibitem[\protect\citeauthoryear{{The DES Collaboration} et~al.,}{{The DES
  Collaboration} et~al.}{2015}]{DES}
{The DES Collaboration} et~al., 2015, preprint, \href
  {http://adsabs.harvard.edu/abs/2015arXiv150302584T} {} (\mn@eprint {arXiv}
  {1503.02584})

\bibitem[\protect\citeauthoryear{{Tilvi} et~al.,}{{Tilvi}
  et~al.}{2013}]{Tilvi13}
{Tilvi} V.,  et~al., 2013, preprint, \href
  {http://adsabs.harvard.edu/abs/2013arXiv1304.4227T} {} (\mn@eprint {arXiv}
  {1304.4227})

\bibitem[\protect\citeauthoryear{{Trenti} \& {Stiavelli}}{{Trenti} \&
  {Stiavelli}}{2008}]{Trenti08}
{Trenti} M.,  {Stiavelli} M.,  2008, \mn@doi [\apj] {10.1086/528674}, \href
  {http://adsabs.harvard.edu/abs/2008ApJ...676..767T} {676, 767}

\bibitem[\protect\citeauthoryear{{Trenti} et~al.,}{{Trenti}
  et~al.}{2011}]{Trenti11}
{Trenti} M.,  et~al., 2011, \mn@doi [\apjl] {10.1088/2041-8205/727/2/L39},
  \href {http://adsabs.harvard.edu/abs/2011ApJ...727L..39T} {727, L39}

\bibitem[\protect\citeauthoryear{{Trotta}}{{Trotta}}{2008}]{Trotta}
{Trotta} R.,  2008, \mn@doi [Contemporary Physics] {10.1080/00107510802066753},
  \href {http://adsabs.harvard.edu/abs/2008ConPh..49...71T} {49, 71}

\bibitem[\protect\citeauthoryear{{Weidner}, {Kroupa}  \&
  {Pflamm-Altenburg}}{{Weidner} et~al.}{2013}]{Weidner13}
{Weidner} C.,  {Kroupa} P.,   {Pflamm-Altenburg} J.,  2013, \mn@doi [\mnras]
  {10.1093/mnras/stt1002}, \href
  {http://adsabs.harvard.edu/abs/2013MNRAS.434...84W} {434, 84}

\bibitem[\protect\citeauthoryear{{Wilkins}, {Stanway}  \& {Bremer}}{{Wilkins}
  et~al.}{2014}]{Wilkins14}
{Wilkins} S.~M.,  {Stanway} E.~R.,   {Bremer} M.~N.,  2014, \mn@doi [\mnras]
  {10.1093/mnras/stu029}, \href
  {http://adsabs.harvard.edu/abs/2014MNRAS.439.1038W} {439, 1038}

\bibitem[\protect\citeauthoryear{{Yoshii}}{{Yoshii}}{1982}]{Yoshii82}
{Yoshii} Y.,  1982, \pasj, \href
  {http://adsabs.harvard.edu/abs/1982PASJ...34..365Y} {34, 365}

\bibitem[\protect\citeauthoryear{{Zheng}, {Flynn}, {Gould}, {Bahcall}  \&
  {Salim}}{{Zheng} et~al.}{2001}]{Zheng01}
{Zheng} Z.,  {Flynn} C.,  {Gould} A.,  {Bahcall} J.~N.,   {Salim} S.,  2001,
  \mn@doi [\apj] {10.1086/321485}, \href
  {http://adsabs.harvard.edu/abs/2001ApJ...555..393Z} {555, 393}

\bibitem[\protect\citeauthoryear{{van Dokkum}, {Abraham}  \& {Merritt}}{{van
  Dokkum} et~al.}{2014}]{van-Dokkum14}
{van Dokkum} P.~G.,  {Abraham} R.,   {Merritt} A.,  2014, \mn@doi [\apjl]
  {10.1088/2041-8205/782/2/L24}, \href
  {http://adsabs.harvard.edu/abs/2014ApJ...782L..24V} {782, L24}

\bibitem[\protect\citeauthoryear{{van der Kruit} \& {Searle}}{{van der Kruit}
  \& {Searle}}{1981a}]{van-der-Kruit81}
{van der Kruit} P.~C.,  {Searle} L.,  1981a, \aap, \href
  {http://adsabs.harvard.edu/abs/1981A%26A....95..105V} {95, 105}

\bibitem[\protect\citeauthoryear{{van der Kruit} \& {Searle}}{{van der Kruit}
  \& {Searle}}{1981b}]{vdKruit81a}
{van der Kruit} P.~C.,  {Searle} L.,  1981b, \aap, \href
  {http://adsabs.harvard.edu/cgi-bin/nph-bib_query?bibcode=1981A%26A....95..10%
5V&db_key=AST} {95, 105}

\makeatother
\end{thebibliography}


\begin{landscape}

\begin{table*}
{\tiny
\caption{The updated values of the M-dwarf found in BoRG with correct Galactic Coordinates. Full machine-readable table will be available in the electronic version of the manuscript.}
\begin{center}
\begin{tabular}{l l l l l l l l l l l l l l l l l l l l l l}
\hline

ID   & ra &  dec &   b  &  l &  $m_{f098w}$ &   $m_{f125w}$ &   $m_{f125w}$ &  $m_{f606w}$ &   M-type &  Mod. &  dist &  Height &  Radius & d-err & Volume & Area \\ 
 & (deg) & (deg) & (deg) & (deg) & (mag) & (mag) & (mag) & (mag) & & & (kpc) & (kpc) & (kpc) & & & arcmin$^2$ \\
\hline
\hline
borg\_0110-0224.551.0 &  17.501759 & -2.418265 & 133.918256 & -64.893095 & 23.82 & 0.04 & 23.46 & 0.02 & 23.28 & 0.03 & 25.17 & 0.09 & 2.04 & 0.32 & 16.48 & 19.74 & 14.25 & 16.88 & 1.85 & 228.43 & 10.88 \\ 
borg\_0110-0224.719.0 &  17.503586 & -2.415663 & 133.921454 & -64.890220 & 24.33 & 0.05 & 23.97 & 0.03 & 23.87 & 0.04 & 25.83 & 0.14 & 2.56 & 0.50 & 16.73 & 22.16 & 15.99 & 17.90 & 2.73 & 287.73 & 10.88 \\ 
borg\_0110-0224.820.0 &  17.528020 & -2.411389 & 133.976421 & -64.881854 & 19.68 & 0.00 & 19.51 & 0.00 & 19.45 & 0.00 & 22.77 & 0.01 & 7.30 & 0.04 & 8.59 & 0.52 & 0.40 & 8.72 & 0.01 & 0.16 & 10.88 \\ 
borg\_0110-0224.1016.0 &  17.553100 & -2.408888 & 134.033565 & -64.875102 & 24.54 & 0.09 & 23.90 & 0.04 & 23.72 & 0.05 & 25.39 & 0.13 & 1.30 & 0.46 & 17.18 & 27.26 & 19.63 & 20.08 & 3.83 & 435.66 & 10.88 \\ 
borg\_0110-0224.1414.0 &  17.559120 & -2.400783 & 134.044244 & -64.866085 & 24.08 & 0.06 & 23.71 & 0.06 & 23.44 & 0.04 & 24.40 & 0.05 & -1.43 & 0.27 & 17.26 & 28.31 & 20.37 & 20.52 & 687.31 & 469.66 & 10.88 \\ 

\hline
\hline
\end{tabular}
\end{center}
\label{t:Mdwarfs}
}
\end{table*}%

\end{landscape}

\end{document}